\documentclass[journal]{IEEEtran}

\usepackage{graphics} % for pdf, bitmapped graphics files
\usepackage{epsfig} % for postscript graphics files
\usepackage{amssymb} % assumes amsmath package installed
\usepackage[cmex10]{amsmath}
\usepackage{cite}
\usepackage{tikz}
\usepackage{caption}
\usepackage{subcaption}

\usepackage{url}
\usepackage{bm}
\usepackage{mathtools}
\usepackage{marvosym}
\usepackage{textcomp}
\usepackage{gensymb}
\usepackage{epstopdf, tikz, environ, float}
\usetikzlibrary{shapes,arrows,calc,patterns, decorations.pathreplacing, backgrounds,positioning,fit}

\newtheorem{theorem}{Theorem}
\newtheorem{lemma}{Lemma}

\newtheorem{assumption}{Assumption}

\newtheorem{example}{Example}
\newtheorem{remark}{Remark}
\newtheorem{proposition}{Proposition}

\DeclareMathOperator{\id}{Id}
\DeclareMathOperator{\sat}{sat}

\DeclareMathOperator{\card}{card}
\DeclareMathOperator{\esssup}{ess~sup}

\DeclareMathOperator{\diag}{diag}

\newcommand{\argmin}{\operatornamewithlimits{argmin}}
\newcommand{\argmax}{\operatornamewithlimits{argmax}}

\begin{document}

\title{Quadratic Programming for Continuous Control of Safety-Critical Multi-Agent Systems Under Uncertainty}
\author{Si Wu, Tengfei Liu, \IEEEmembership{Senior Member, IEEE}, Magnus Egerstedt, \IEEEmembership{Fellow, IEEE} and Zhong-Ping Jiang, \IEEEmembership{Fellow, IEEE}% <-this % stops a space
\thanks{This work was supported in part by NSFC grants 61633007 and 61533007, and in part by NSF grant EPCN-1903781.}% <-this % stops a space
\thanks{S. Wu is with State Key Laboratory of Synthetical Automation for Process Industries, Northeastern University, Shenyang, 110004, China {\tt\small (e-mail: wusixstx@163.com; 2010306@stu.neu.edu.cn)}}%
\thanks{T. Liu is with State Key Laboratory of Synthetical Automation for Process Industries, Northeastern University, Shenyang, 110004, China {\tt\small (e-mail: tfliu@mail.neu.edu.cn)}}%
\thanks{M. Egerstedt is with the Samueli School of Engineering, University of California, Irvine, CA 92697, USA {\tt\small (e-mail:magnus@uci.edu)}}%
\thanks{Z. P. Jiang is with Department of Electrical and Computer Engineering, New York University, 360 Jay Street, Brooklyn, NY 11201, USA {\tt\small (e-mail: zjiang@nyu.edu)}}%
\thanks{Corresponding author: T. Liu.}
}
\maketitle

\begin{abstract}
This paper studies the control problem for safety-critical multi-agent systems based on quadratic programming (QP). Each controlled agent is modeled as a cascade connection of an integrator and an uncertain nonlinear actuation system. In particular, the integrator represents the position-velocity relation, and the actuation system describes the dynamic response of the actual velocity to the velocity reference signal.
The notion of input-to-output stability (IOS) is employed to characterize the essential velocity-tracking capability of the actuation system. The uncertain actuation dynamics may cause infeasibility or discontinuous solutions of QP algorithms for collision avoidance. Also, the interaction between the controlled integrator and the uncertain actuation dynamics may lead to significant robustness issues.
By using nonlinear control methods and numerical optimization methods, this paper first contributes a new feasible-set reshaping technique and a refined QP algorithm for feasibility, robustness, and local Lipschitz continuity. Then, we present a nonlinear small-gain analysis to handle the inherent interaction for guaranteed safety of the closed-loop multi-agent system. 
The proposed methods are illustrated by numerical simulations and a physical experiment.
\end{abstract}


\section{Introduction}
\label{section.introduction}

Attaining primary objectives while satisfying motion constraints is an essential yet challenging task for vehicles and robotic systems. This has been one of the most attractive topics in the interdisciplinary literature of controls and robotics in the past decades \cite{Latombe-book-1991,Arkin-book-1998,Choset-book-2005,Ren-Beard-book-2008,Bullo-Cortes-Martinez-book-2009,Mesbahi-Egerstedt-book-2010}.

Early utilized in constrained optimization \cite{Boyd-Vandenberghe-book-2004}, barrier functions have been employed to characterize state constraints for nonlinear control systems. Control barrier functions have been developed to enable constrained control designs \cite{Polak-Yang-Mayne-SIAMControl-1993,Wills-Heath-Auto-2004,Ngo-Mahony-Jiang-CDC-2005,Tee-Ge-Tay-Auto-2009,Wieland-Allgower-NOLCOS-2007,Prajna-Jadbabaie-Pappas-TAC-2007,Ames-Grizzle-Tabuada-CDC-2014,Wisniewski-Sloth-TAC-2016,Romdlony-Jayawardhana-Auto-2016}. The substantial relationship between control barrier functions and control Lyapunov functions \cite{Artstein-NA-1983} opens the door to a systematic development of a multi-objective control theory \cite{Ames-Xu-Grizzle-Tabuada-TAC-2017}. In particular, the recent advancement of control barrier functions relaxes the requirement of invariance of every level set (see \cite{Xu-Tabuada-Grizzle-Ames-IFAC-2015} for zeroing barrier functions). Still, it only assumes an increasing property of the barrier function when the system state is outside the desired safety set. Moreover, for a control-affine nonlinear system, an appropriately defined control barrier function entails linear inequality constraints on admissible control inputs to keep the system state inside the desired set. See \cite{Jankovic-Auto-2018} for a discussion on the half-space robustness property of Sontag's formula \cite{Sontag-SCL-1989}, and pointwise minimum norm formula \cite{Freeman-Kokotovic-book-1996}. This treatment allows computationally efficient integration of different control strategies to fulfill conflicting constraints.

Indeed, quadratic programming (QP) is a powerful tool for real-time synthesis of controllers by incorporating different specifications simultaneously \cite{Escande-Mansard-Wieber-IJRR-2014,Mellinger-Kumar-ICRA-2011,Ames-Powell-CPS-2013}. For a system subject to motion constraints, a QP algorithm calculates the admissible control input that fulfills the constraints and is as close as possible to the set of control inputs for primary objectives \cite{Ames-Xu-Grizzle-Tabuada-TAC-2017}.
The notions of robust barrier functions \cite{Ames-Xu-Grizzle-Tabuada-TAC-2017,Jankovic-Auto-2018} and input-to-state safety \cite{Romdlony-Jayawardhana-CDC-2016,Kolathaya-Ames-CSL-2019} have been developed to handle perturbations.
The study of Lipschitz continuity of QP-based control laws is not only of theoretical interest for well-defined solutions of closed-loop systems but also beneficial to avoiding chattering and other unexpected transient behaviors in practice \cite{Ong-Cortes-CDC-2019, Morris-Powell-Ames-CDC-2015,Ames-Xu-Grizzle-Tabuada-TAC-2017,Jankovic-Auto-2018}.
The integration of barrier functions and QP algorithms have found various applications, including automotive safety \cite{Ames-Xu-Grizzle-Tabuada-TAC-2017}, robotic locomotion and manipulation \cite{Morris-Powell-Ames-CDC-2015,Cortez-Oetomo-Manzie-Choong-TCST-2019}, multi-robot systems \cite{Wang-Ames-Egerstedt-TRO-2017,Glotfelter-Cortes-Egerstedt-CSL-2017,Wilson-Egerstedt-CSM-2020}. See \cite{Ames-Coogan-Egerstedt-ECC-2019} for a recent survey on control barrier functions and QP-based controller synthesis.

Other velocity obstacle approach \cite{Fiorini-Shiller-IJRR-1998,Berg-Guy-Lin-Manocha11-RR-2011} calculates the set of feasible velocities for constrained motions. Still, the underlying assumption of piecewise continuous speed or acceleration possibly results in limited robustness to dynamic uncertainties.
We also recognize the refined designs to address the discontinuity issue \cite{Rufli-Alonso-Mora-Siegwart-TRO-2013}.
Interested readers may also consult the recent paper \cite{Singletary-Klingebiel-Bourne-Ames-IROS-2021} for a comparative study of control barrier functions and the popular artificial potential fields \cite{Khatib-JRR-1986}.

\begin{figure}[h!]
\centering
\includegraphics[width=\linewidth]{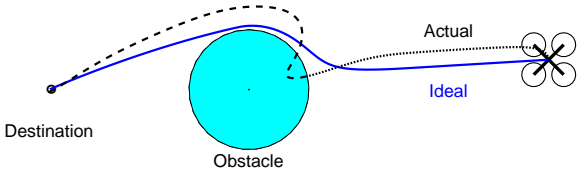}
\caption{The ideal path is collision-free, while uncertain actuation dynamics may cause collision.}
\label{figure.collisionfree}
\end{figure}

This paper investigates the safety control problem for a class of mobile agents modeled as a cascade connection of an integrator and an uncertain actuation system. Such a system setup covers a broad class of practical control systems. 
If the dynamics of the actuation system are neglectable, then the proposed model is reduced to an integrator, which is known as an essential model for safety control. Some other systems, e.g., double-integrators and Euler-Lagrange systems, can also be transformed into our model by introducing appropriate virtual control laws. Interestingly, the identified model of a quadrotor is in the form of our model (see Section \ref{section.experiment}).
Unsurprisingly, dynamic uncertainties challenge the robustness and computational feasibility of QP-based algorithms and may cause the collision avoidance performance to deteriorate (as illustrated in Figure \ref{figure.collisionfree}).

%{\color{red} \bf
%    The uncertain actuation system is consist of the dynamic model of the agent and its controller.
%    
%    We call the dynamic model of agent and its controller actuation.
%    For the dynamic model, there are many researches on the possible time-varying, nonlinear, uncertainty and disturbance in the dynamic model. And the controller they proposed can make the actuation have tracking ability.
%    Considering the dynamic model in the design process of safety controller will make the safety controller more complex and bring difficulties to the implementation of algorithm.
%
%    
%Motivated by hierarchical control, considering the model of actuation in the design process of safety controller will make the safety controller more complex and bring difficulties to the implementation of algorithm.
%The actuation is the result of the closed-loop model of the controller and the dynamic model of agent. 
%The actuation may be time-varying and unknown, but it has tracking ability.
%Considering the above reasons, we abstracts the actuation into a system with input-to-output property stability.
%}

This paper assumes an essential velocity-tracking capability for the actuation system, which is described by the notion of input-to-output stability (IOS) \cite{Jiang-Teel-Praly-MCSS-1994,Sontag-2008}. In the presence of uncertain actuation dynamics, a conventional QP algorithm may not be feasible, and the solution may be discontinuous even if it exists. Moreover, the uncertain actuation system leads to an unexpected feedback loop from the constrained position to the velocity-tracking error and possibly destroys the usual half-space robustness.

This paper proposes a seamless integration of numerical optimization and nonlinear control to address the major technical difficulty caused by uncertain actuation dynamics. Our first contribution lies in a new feasible-set reshaping technique by refining a standard QP algorithm for guaranteed feasibility, robustness, and local Lipschitz continuity. Based on the treatment above, the controlled multi-agent system is transformed into an interconnected system composed of two subsystems, one corresponding to the nominal controlled system subject to the velocity-tracking error and the other caused by the uncertain actuation dynamics. We employ gains to represent the interconnections and develop a nonlinear small-gain analysis to guarantee safety.

To the best of our knowledge, some techniques in this paper are reported for the first time. The feasible-set reshaping technique ensures (local) Lipschitz continuity of the solution of the refined QP algorithm and would be beneficial to other related problems with continuous motion constraints. The robustness analysis for collision avoidance subject to multiple safety constraints is still valuable when the controlled agents are free of uncertain actuation dynamics. The small-gain analysis takes advantage of the inherent interaction between the nominal system and the uncertain actuation dynamics. It motivates a new integration methodology for kinematics and dynamics control loops.

The rest of the paper is organized as follows. Section \ref{section.problemformulation} introduces the system setup and gives the collision avoidance problem formulation. In Section  \ref{section.motivatingexamples}, we employ two examples to discuss the technical difficulty caused by the uncertain actuation dynamics. The main result of a refined QP algorithm with a reshaped feasible set is presented in Section \ref{section.reshapefeasibleset}. The proof of the main result, given in Section \ref{section.proofs}, is based on several new properties of the refined QP algorithm and local small-gain analysis. In Section \ref{section.experiment}, we employ numerical simulations and a physical experiment based on quadrotors to illustrate the validity of the proposed method. Section \ref{section.conclusion} presents some concluding remarks. Due to space limitations, some proofs are placed in the technical report \cite{technicalreport}.

\subsection*{Notations}

The following notations are given to make the paper self-contained.

% matrix
We use $|x|$ to represent the Euclidean norm of $x\in\mathbb{R}^n$, and use $|A|$ to represent the induced $2$-norm of $A\in\mathbb{R}^{m\times n}$. 
For a nonzero real vector $x$, we denote $\hat{x} = x/|x|$. For vectors $x,x'\in\mathbb{R}^n$, $x<x'$ represents that the corresponding elements $x_i$ and $x_i'$ satisfy $x_i<x_i'$. The notations $\leq$, $>$ and $\geq$ are defined in the same way for vectors. For a real vector $x$, $\min\{x\}$ and $\max\{x\}$ denote the smallest and the largest element of $x$, respectively. 
For a real vector $x$, $[x]_i$ represents the $i$-th element.
For a real matrix $A$, $[A]_{i,j}$, $[A]_{i,:}$ and $[A]_{:, j}$ represent the element at the $i$-th row and the $j$-th column, the $i$-th row vector and the $j$-th column vector of the matrix $A$, respectively, $\min\{A\}$ denotes a column vector containing the minimum value of each row, and $\max\{A\}$ denotes a column vector containing the maximum value of each row. We use $\otimes$ to represent the Kronecker product, and use $\odot$ to represent the Hadamard product. In particular, for $A\in\mathbb{R}^{m\times n}$ and $B\in\mathbb{R}^{1\times n}$, $C = A\odot B$ is defined by $[C]_{i,j}=[A]_{i,j}[B]_{1,j}$. For $A\in\mathbb{R}^{n\times n}$, $\lambda_{\max}(A)$ and $\lambda_{\min}(A)$ take the maximum eigenvalue and the minimum eigenvalue, respectively, and $\sigma_{\max}(A)$ and $\sigma_{\min}(A)$ take the maximum singular value and the minimum singular value, respectively.

% signal
For a measurable and locally essentially bounded signal $u:\mathbb{R}_+\rightarrow\mathbb{R}^m$, $\|u\|_{[t_1, t_2)}=\esssup_{\tau\in[t_1,t_2)}|u(\tau)|$, and $\|u\|_t=\|u\|_{[0, t)}$.
Let $f(t)$ be a real-valued function defined over an open interval $(a,b)$. The upper right Dini derivative of $f(t)$ at $t_0\in (a, b)$ is defined as $D^+f(t_0)= \limsup_{t\rightarrow t_{0^+}} (f(t)-f(t_0))/(t-t_0)$.

% function
The definitions of positive definite functions and functions of classes $\mathcal{K}$, $\mathcal{K}_{\infty}$ and $\mathcal{KL}$ can be found in \cite{Khalil-book-2002}. A continuous function $\alpha:(-a,b)\rightarrow(-c,\infty)$ with constants $a,b,c>0$ is said to be of class $\mathcal{K}^e$ if it is strictly increasing and $\alpha(0)=0$. A continuously differentiable function $\mu:(-a,\infty)\rightarrow(0,b)$ with constants $a$, $b>0$ is said to be of class $\mathcal{M^-C}$, denoted by $\mu\in\mathcal{M^-C}$, if it is strictly decreasing, strictly convex, $\lim_{s\rightarrow\infty}\mu(s)=0$, and $\lim_{s\rightarrow \infty}\partial\mu(s)/\partial s=0$. $\id$ denotes the identity function. $\sat(r):=\max\{\min\{r,1\},0\}$ is a saturation function defined for $r\in\mathbb{R}$.

\section{Problem Formulation and Preliminaries}
\label{section.problemformulation}

%This section first introduces the class of systems to be studied in this paper, 
%{\color{red} \bf gives the problem formulation of collision avoidance, proposes the assumptions on actuation system and velocity command, and finally gives the characterization of safety}.

This section first introduces the class of systems to be studied in this paper, and then gives the problem formulation of collision avoidance.
%{\color{red}\bf
%After that, we make the assumptions on actuation system and velocity command.
%Finally, the characterization of safety is given in the end of this section.
%}

Suppose there are $n_a$ agents and $n_o$ obstacles indexed by 
\begin{align}
\mathcal{N}_a =\{1,\ldots,n_a\}, ~~~\mathcal{N}_o =\{n_a + 1,\ldots, n_a + n_o\},\label{eq.def.NaNo}
\end{align}
respectively.
Denote $\mathcal{N}_{ao}=\mathcal{N}_a\cup\mathcal{N}_o$, and $n_{ao}=n_a+n_o$. For each $i\in\mathcal{N}_{ao}$, we use $p_i\in\mathbb{R}^n$ to represent the position of the mobile agent or the obstacle, and use
\begin{align}
p =\left[p_1,\ldots,p_{n_{ao}} \right] \label{eq.def.bmP}
\end{align}
to represent the positions of all the mobile agents and the obstacles.

For $i\in\mathcal{N}_a$, each agent $i$ is modeled as a cascade connection of a nominal system and an uncertain actuation system. Specifically, the nominal system is represented by
\begin{align}
\dot{p}_i=v_i,\label{eq.def.p_i.kinematics}
\end{align}
where $v_i\in\mathbb{R}^n$ is the velocity. The velocity $v_i$ is generated by an actuation system, which is in the general form:
\begin{align}
\dot{z}_i=f(z_i,v^*_i),~~v_i=g(z_i,v^*_i)\label{eq.def.v_i.actuation}
\end{align}
where $z_i\in\mathbb{R}^m$ is the state of the actuation system, $v^*_i\in\mathbb{R}^n$ represents the velocity reference signal, and $f$ and $g$ are locally Lipschitz functions. 

%We use $v_i^c\in\mathbb{R}^n$ to represent the velocity command signal for the achievement of primary control objective.
%%
%Our objective is to design controllers that incorporate velocity command signals and safety constraints such that if the agents satisfy specific initial conditions, then
%\begin{align}
%\min_{i\in\mathcal{N}_a,~j\in\mathcal{N}_{ao}\setminus\{i\}}|p_i(t)-p_j(t)|\ge D\label{eq.problemformulation}
%\end{align}
%for all $t\geq 0$, where $D$ is a positive constant {\color{red} \bf representing safe distance}.
%The expected control system is shown in Figure \ref{figure.systemdiagram}. {\color{red} \bf With such objective, the following definition is presented to describe the safety of the multi-agent system.}

%The multi-agent system is said to be safe if under same specific initial condition, the position of the agents satisfy
%\begin{align}
%    \min_{i\in\mathcal{N}_a,~j\in\mathcal{N}_{ao}\setminus\{i\}}|p_i(t)-p_j(t)|\ge D\label{eq.problemformulation}
%\end{align}
%The multi-agent system indexed by $\mathcal{N}_a$, with dynamics given in \eqref{eq.def.p_i.kinematics} and \eqref{eq.def.v_i.actuation}, is said to be safe if under some specific initial conditions 

The multi-agent system with each agent described by \eqref{eq.def.p_i.kinematics}-\eqref{eq.def.v_i.actuation} is said to be safe if under some specific initial conditions 
\begin{align}
\min_{i\in\mathcal{N}_a,~j\in\mathcal{N}_{ao}\setminus\{i\}}|p_i(0)-p_j(0)|\ge \underline{D}_0, ~~\max_{i\in\mathcal{N}_a} |z_i| \le \bar{z}_0, \label{eq.problemformulation.initialcondition}
\end{align}
with $\underline{D}_0$, $\bar{z}_0>0$, the relative positions of the multi-agent system satisfy
\begin{align}
\min_{i\in\mathcal{N}_a,~j\in\mathcal{N}_{ao}\setminus\{i\}}|p_i(t)-p_j(t)|\ge D, \label{eq.problemformulation}
\end{align}
for all $t\ge 0$, where $D>0$ is the safety distance.

\begin{figure}[h!]
\centering
\begin{tikzpicture}[scale=0.77]
% collision avoidance controller
\node at (-3.9,-0.7) {Safety};
\node at (-3.9,-1.5) {Controller};
\draw (-6,0) rectangle (-1.8,-2.2);

% actuation
\node at (-3.9,2.2) {Actuation System};
\node at (0.9,2.2) {Nominal System};
\draw (-6,2.8) rectangle (-1.8,0.9);
\node at (-3.9,1.4) {\eqref{eq.def.v_i.actuation}};

% nominal system
\node at (0.9,1.4) {\eqref{eq.def.p_i.kinematics}};
\draw (-9mm,9mm) rectangle (27mm,28mm);

% positions-i
\node at (3.7,2.1) {$p_i$};
\draw[-latex] (3.3,1.8) -- (4,1.8);
\draw[-latex] (2.7,1.8) -- (3.3,1.8) -- (3.3,-0.5)  -- (-1.8,-0.5);
\draw[fill=black] (3.3,1.8) circle (0.03);

% positions-j
\node[right] at (1.2,-1.7) {$p_j$ ($j\neq i$)};
\draw[-latex] (1.1,-1.7) -- (-1.8,-1.7);

% velocity command
\node at (-3.6,0.3) {$v^*_i$};
\draw[-latex] (-3.9,0) -- (-3.9,0.9);

% velocity
\node at (-1.4,2.1) {$v_i$};
\draw[-latex] (-1.8,1.8) -- (-0.9,1.8);

% ideal velocity
\node[right] at (-7.2,-0.8) {$v^c_i$};
\draw[-latex] (-7,-1.1) -- (-6,-1.1);

\node at (1.3,3.3) {Mobile Agent};
\draw[dashed]  (-6.2,3) rectangle (2.9,0.7);
\end{tikzpicture}
\caption{Block diagram of the safety control system.}
\label{figure.systemdiagram}
\end{figure}
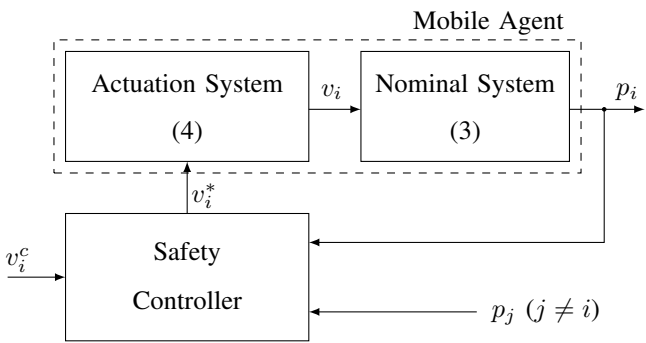

We use $v_i^c\in\mathbb{R}^n$ to represent the velocity command signal for primary control objective.
The safety control problem is concerned about designing controllers for the agents to incorporate the velocity command signals and the safety qualifications.
%Our objective is to design controllers that incorporate velocity command signals and safety qualifications.
% such that if the agents satisfy specific initial conditions, then
The expected control system is shown in Figure \ref{figure.systemdiagram}.

%{\color{red} \bf
%\begin{definition}[Safety] \label{definition.safety}
%The multi-agent system indexed by $\mathcal{N}_{a}$, with dynamics given in \eqref{eq.def.p_i.kinematics} and \eqref{eq.def.v_i.actuation}, is safe if the relative position satisfies \eqref{eq.problemformulation} for all $t\ge 0$.
%\end{definition}   
%}
%

\subsection{Assumptions on the Actuation Systems and the Velocity Command Signals}

In the presence of the uncertain actuation dynamics, the actual velocity $v_i$ may not equal to the velocity reference signal $v^*_i$. Define velocity-tracking error
\begin{align}
\tilde{v}_i=v_i-v^*_i.\label{eq.multi.referencetrackingerror}
\end{align}
Then, the nominal system \eqref{eq.def.p_i.kinematics} can be rewritten as
\begin{align}
\dot{p}_i&=v_i=v^*_i+\tilde{v}_i.\label{eq.multi.kinematicstransform}
\end{align}
In practice, it is essential that the actuation system is inherently stable and admits some velocity-tracking capability. In this paper, we assume that for a constant reference signal $v^*_i$, the actual velocity $v_i$ asymptotically converges to $v^*_i$, and for a time-varying $v^*_i$, the tracking error $\tilde{v}_i$ depends on the changing rate of $v^*_i$.

Suppose that the velocity reference signal $v^*_i$ is locally Lipschitz on the time-line, and denote
\begin{align}
v^{*d}_i(t)=D^+v^*_i(t).\label{eq.def.v*d_i.derivative}
\end{align}
The following assumption is made on the uncertain actuation system.

\begin{assumption}[Stability and reference-tracking capability of the actuation system]
\label{assumption.multi.referencetracking}
Consider the actuation system \eqref{eq.def.v_i.actuation}. There exist locally Lipschitz functions $\alpha_{z1},\alpha_{z2},\gamma^{v^{*d}}_{\tilde{v}}\in\mathcal{K}$ and constants $c_{\tilde{v}} \geq 1$ and $\lambda>0$ such that for any $z_i(0)$ and any locally Lipschitz and bounded $v_i^*$,
\begin{align}
|\tilde{v}_i(t)|&\leq \beta_{\tilde{v}}(|z_i(0)|,t) + \gamma^{v^{*d}}_{\tilde{v}}(\|v^{*d}_i\|_{t})\label{eq.asmp.vtilde_i.IOpS}\\
|z_i(t)|&\le\alpha_{z1}(|z_i(0)|)+\alpha_{z2}(\|v^*_i\|_{t}) \label{eq.asmp.z_i.ISS}
\end{align}
hold for all $t\geq 0$, where $\beta_{\tilde{v}}(s,t)=c_{\tilde{v}} e^{-\lambda t}s$ for $s,t\in\mathbb{R}_+$.
\end{assumption}

\begin{remark} % describe the properties of actuation uncertainty
With the velocity-reference signal $v^*_i$'s Dini derivative as the input and the velocity-tracking error $\tilde{v}_i$ as the output, property \eqref{eq.asmp.vtilde_i.IOpS} employs the notion of IOS to characterize the velocity-tracking capability: $\tilde{v}_i$ ultimately converges to the range of $\gamma^{v^{*d}}_{\tilde{v}}(\|v^{*d}\|_{\infty})$. Given bounded velocity reference signal $v^*_i$, property \eqref{eq.asmp.z_i.ISS} guarantees the boundedness of state $z_i$. See, e.g., \cite{Sontag-2008,Liu-Jiang-Hill-book-2014,Jiang-Liu-ARC-2018} for tutorials of the related notions. 
% The basic idea of this paper is still valid when the agents' actuation systems are different but admit the stability and the reference tracking capabilities as given by Assumption \ref{assumption.multi.referencetracking}.
\end{remark}

\begin{remark} % universality of the assumption
%The models of many types of agents could satisfy Assumption \ref{assumption.multi.referencetracking} by using appropriate transformations.
%The differential-drive robot could transforming into single-integrator model \cite{Glotfelter-Buckley-Magnus-RAL-2019,Olfati-Saber-ACC-2002}.
%The integrators system studied in \cite{Wang-Ames-Egerstedt-TRO-2017} could satisfy the assumption by designing a virtual velocity controller.
%And some specific affine systems can be transformed into integrators with corresponding orders\cite{Khalil-book-2002, Morris-Powell-Ames-CDC-2015}. 
%
Suppose the dynamics of the actuation system is neglectable. In that case, i.e., $v^*_i\equiv v_i$, the agent model is reduced to a single-integrator, which has been widely studied in the literature of safety control \cite{Glotfelter-Buckley-Magnus-RAL-2019}.
The double-integrator model \cite{Wang-Ames-Egerstedt-TRO-2017} can also be transformed into the form of \eqref{eq.def.p_i.kinematics}-\eqref{eq.def.v_i.actuation} by introducing a virtual control law to the velocity loop.
Interestingly, the experimental data of a quadrotor in Section \ref{section.experiment} also coincides with model \eqref{eq.def.p_i.kinematics}-\eqref{eq.def.v_i.actuation} and satisfies Assumption \ref{assumption.multi.referencetracking}.
\end{remark}

Example \ref{example.TypeKsystem} gives a condition for linear control systems to satisfy Assumption \ref{assumption.multi.referencetracking}. 
%Interestingly, Assumption \ref{assumption.multi.referencetracking} is also in accordance with the experimental data of a quadrotor in Section \ref{section.experiment}.

\begin{example}[Reference tracking capability of linear control systems]
\label{example.TypeKsystem}
Consider a linear control system
\begin{align}
\dot{z} = Az + Bv^*, ~~ v=Cz \label{eq.example.linearsystem}
\end{align}
where $z\in\mathbb{R}^m$ is the state, $v\in\mathbb{R}^n$ is the output, $v^*\in\mathbb{R}^n$ is the input, and $A$, $B$ and $C$ are real matrices with appropriate dimensions. It is assumed that $A$ is a Hurwitz matrix and satisfies
\begin{align}
CA^{-1}B = -I_{n\times n}. \label{eq.example.linearsystem.condition}
\end{align}
Define the tracking error $\tilde{v}=v-v^*$. For any constant input $v^*$, condition \eqref{eq.example.linearsystem.condition} implies that $\lim_{t\rightarrow\infty}\tilde{v}(t)=0$.

Suppose that $v^*$ is continuously differentiable, and denote $v^{*d}(t)=\dot{v}^*(t)$. Then, define $\zeta = Az+Bv^*$. Direct calculation yields:
\begin{align}
\dot{\zeta}&=A\dot{z}+Bv^{*d}=A\zeta+B v^{*d},\label{eq.example.tranformed.linearsystem1}\\
\tilde{v}&=Cz-v^* =CA^{-1}(\zeta-Bv^*)-v^*=CA^{-1}\zeta.\label{eq.example.tranformed.linearsystem2}
\end{align}
Because $A$ is Hurwitz, the transformed linear control system \eqref{eq.example.tranformed.linearsystem1}--\eqref{eq.example.tranformed.linearsystem2} is stable. Property \eqref{eq.asmp.vtilde_i.IOpS} can be proved by directly applying the definition of IOS-Lyapunov function \cite{Sontag-Wang-SIAMControl-2000}; see Section \ref{subsection.identification} for details. 
\hfill $\Diamond$
\end{example}

In practice, $v^c_i$ is generated by the primary controller. Without loss of generality, we make the following assumption on the velocity command signal $v_i^c$.

\begin{assumption}[Boundedness of the velocity command and its derivative]
\label{assumption.multi.boundedprimarycontrol}
For each $i\in\mathcal{N}_a$, $v_i^c$ is continuously differentiable with respect to time, and there exist positive constants $\bar{v}^c$ and $\bar{v}^{cd}$ such that
\begin{align}
|v^c_i(t)|\le\bar{v}^c,~~|\dot{v}^c_i(t)|\le\bar{v}^{cd}
\end{align}
for all $t\ge0$.
\end{assumption}

\subsection{Characterization of Safety}

For the multi-agent system \eqref{eq.def.p_i.kinematics}--\eqref{eq.def.v_i.actuation}, define
\begin{align}
\tilde{p}_{ij}=p_i-p_j\label{eq.multi.kinematicstransform2}
\end{align}
to represent the relative positions. 
Then, from \eqref{eq.multi.kinematicstransform}, we have 
\begin{align}
\dot{\tilde{p}}_{ij} = v_i - v_j = v^*_i-v^*_j + \tilde{v}_i-\tilde{v}_j. \label{eq.model.tildep_ij}
\end{align}

To describe the safety of agent $i$ with respect to another agent or an obstacle $j$, we define
\begin{align}
V(\tilde{p}_{ij})=\mu(|\tilde{p}_{ij}|-D_s), \label{eq.multi.lyapunov}
\end{align}
where $\mu:(-D_s,\infty)\rightarrow\mathbb{R}_+$ is an $\mathcal{M^-C}$ function and $D_s\in\mathbb{R}_+$ is safety margin. 
% Recall the safety defined in \ref{definition.safety}.
If the velocity reference signals $v^*_i$ and $v^*_j$ satisfy
\begin{align}
-\hat{\tilde{p}}_{ij}^T (v^*_i-v^*_j) \le -\alpha_V(V(\tilde{p}_{ij}) - \mu(0))
\end{align}
with $\alpha_V\in\mathcal{K}^e$, then along the trajectories of \eqref{eq.model.tildep_ij}, we have
\begin{align}
\nabla V \dot{\tilde{p}}_{ij} \le  \frac{\partial \mu(|\tilde{p}_{ij}|-D_s)}{\partial (|\tilde{p}_{ij}|-D_s)}\left(\alpha_V( V(\tilde{p}_{ij}) - \mu(0) )+\hat{\tilde{p}}_{ij}^T (\tilde{v}_i-\tilde{v}_j)\right),
\end{align}
% and thus we are able to find positive definite function $\theta$ such that 
%and thus there exists a class $\mathcal{K}_{\infty}$ function $\theta$ such that 
%\begin{align}
%&V(\tilde{p}_{ij}) \ge (\id + \theta)\left(\alpha_V^{-1}(|\tilde{v}_i|+|\tilde{v}_j|)+\mu(0)\right) \Rightarrow \notag \\
%&\nabla V\dot{\tilde{p}}_{ij} \le 0.
%%&\nabla V\dot{\tilde{p}}_{ij} \le \frac{\partial \mu(|\tilde{p}_{ij}|-D_s)}{\partial (|\tilde{p}_{ij}|-D_s)}\left(\id-(\id + \theta)^{-1}\right)(V(\tilde{p}_{ij})).
%\end{align}
%If the initial states satisfy $V(\tilde{p}_{ij}(0))\le \mu(D-D_s)$, and the velocity-tracking errors satisfy $\|\tilde{v}_i\|_{\infty}+\|\tilde{v}_j\|_{\infty}\le\alpha_V((\id+\theta)^{-1}(\mu(D-D_s))-\mu(0)) $, then
%\begin{align}
%V(\tilde{p}_{ij}(t)) \le \mu(D-D_s)
%\end{align}
%holds for all $t\ge 0$, which means safety by recalling equation \eqref{eq.problemformulation}.
and thus 
\begin{align}
V(\tilde{p}_{ij}) \ge \left(\alpha_V^{-1}(|\tilde{v}_i|+|\tilde{v}_j|)+\mu(0)\right) \Rightarrow \nabla V\dot{\tilde{p}}_{ij} \le 0.
\end{align}
If the initial states satisfy $V(\tilde{p}_{ij}(0))\le \mu(D-D_s)$, and the velocity-tracking errors satisfy $\|\tilde{v}_i\|_{\infty}+\|\tilde{v}_j\|_{\infty}\le\alpha_V(\mu(D-D_s)-\mu(0)) $, then
\begin{align}
V(\tilde{p}_{ij}(t)) \le \mu(D-D_s)
\end{align}
holds for all $t\ge 0$, which means safety by recalling equation \eqref{eq.problemformulation}.

\begin{remark}\label{remark.uncertainty_caused_infeasible}
If the agents are free of uncertain actuation dynamics, then the safety control problem would be readily solvable by applying barrier-function-based QP designs\cite{Romdlony-Jayawardhana-Auto-2016} and using the idea of forward invariance \cite[Theorem 1]{Ames-Xu-Grizzle-Tabuada-TAC-2017}. However, for the class of multi-agent systems with uncertain actuation dynamics, the velocity-tracking errors may violate the safety constraint; see the discussions in Section \ref{section.motivatingexamples}.
%In our problem setting, the velocity-tracking errors depend on the changing rate of the reference signal, which leads to an interaction between the nominal systems and the uncertain actuation systems. We propose a small-gain synthesis to address this problem; please see Section \ref{section.proofs} for the proof of the main result.
\end{remark}
%{\color{red} \cite{dimarogonas2006feedback, Berg-Guy-Lin-Manocha11-RR-2011} discussed the collision avoidance of multi-agent systems modeled by single integrators.}

% \section{Limitations of the Existing Methods} %   Limitations   of 
\section{Limitations of Standard Designs} %   Limitations   of 
\label{section.motivatingexamples}

This section employs examples to discuss the technical difficulty caused by the uncertain actuation dynamics.
It is shown that the velocity-tracking errors may either lead to infeasibility of conventional QP-based controllers, or result in non-Lipschitz solutions.
A non-Lipschitz solution may cause unexpected transient response of the uncertain actuation system and destroy the safety of the controlled agent.

\subsection{A QP-based Controller with an Extended Safety Margin}
\label{subsection.extendingsafetymargin}

For convenience of notations, denote
\begin{align}
V_{ij}(t)&=V(\tilde{p}_{ij}(t)). \label{eq.V_ij.simplify}
\end{align}
Given the velocity command $v^c_i$ for the primary control objective, inspired by the QP-based control strategy \cite{Wang-Ames-Egerstedt-TRO-2017,Kolathaya-Ames-CSL-2019}, one may choose the actual velocity reference signal $v_i^*$ such that safety is ensured and at the same time $v_i^*$ is as close to the velocity command $v_i^c$ as possible:
\begin{align}
v^{*}_i&=\argmin_{v^{*}_i\in\mathcal{P}^o_i(p)} \frac{1}{2}v^{*T}_iv^{*}_i - v^{cT}_iv^{*}_i\label{eq.def.QP.conventional}
\end{align}
where
\begin{align}
\mathcal{P}_i^o(p)=\left\{v^{*}_i\in\mathbb{R}^n:A^o_i(p)v^{*}_i+a^o_i(p)\le 0\right\}\label{eq.def.QP.conventional.feasibleset}
\end{align}
is the feasible set with
\begin{align}
A^o_i(p)&=[
-\hat{\tilde{p}}_{i1}, \ldots, -\hat{\tilde{p}}_{i(i-1)},
-\hat{\tilde{p}}_{i(i+1)},\ldots,-\hat{\tilde{p}}_{in_{ao}}]^T, \label{eq.QP2.conventional.Ai}\\
a^o_i(p)&=[\alpha_V(V_{i1}-\mu_0), \ldots, \alpha_V(V_{i(i-1)}-\mu_0),\ldots \notag \\ &~~~~~~\alpha_V(V_{i(i+1)}-\mu_0),\ldots, \alpha_V(V_{in_{ao}}-\mu_0)]^T. \label{eq.QP2.conventional.ai}
\end{align}
Here, $\alpha_V$ is a continuously differentiable class $\mathcal{K}^e$ function, and $\mu_0=\mu(0)$. 
To avoid singularity, the algorithm requires $\tilde{p}_{ij}\neq0$ for $i\in\mathcal{N}_a$ and $j\in\mathcal{N}_{ao}\setminus\{i\}$.

However, the velocity-tracking errors $\tilde{v}_i$ may violate the safety constraint of the conventional QP-based design \cite{Ames-Xu-Grizzle-Tabuada-TAC-2017,Jankovic-Auto-2018}. An intuitive solution is to extend the safety margin \cite{Fox-Burgard-Thrun-RAM-1997}, which, however, still may not guarantee the feasibility of the QP algorithm if the total number of the constraints is larger than one; see Example \ref{example.infeasible}.

\begin{figure}[h!]
\centering
\begin{tikzpicture}[scale=1.5]
% definition of agent
\coordinate [label=above right:$p_1$] (p) at (-1.1889,1.1889) {} {} {};

% velocity command
\draw[-latex] (p) --++(-45:0.45cm) node [near end, below] {$v^c_1$};	

% definition of obstacles
\coordinate (po1) at (0.5,0.5);
\coordinate (po2) at (-0.5,-0.5) {};

% mobile agent
\filldraw[draw=black, fill=cyan!50, scale=0.2, rotate = -45] plot coordinates { (-8.4,0) (-10.2,0.6) (-9.8,0) (-10.2,-0.6) (-8.4,0)};

% Obstacle - 1
\draw [fill=cyan!50, fill opacity = 0.2, dashed, draw=cyan] (po1) circle (1.0);
\draw [fill=cyan!50] (po1) circle (0.6);
\draw [fill=black] (po1) circle (0.02);
\node at (0.5,0.3) {$p_2$};

% Obstacle - 2
\draw [fill=cyan!50, fill opacity = 0.2, dashed, draw=cyan] (po2) circle (1.0);
\draw [fill=cyan!50] (po2) circle (0.6);
\draw [fill=black] (po2) circle (0.02);
\node at (-0.5,-0.7) {$p_3$};

% Trajectory
\draw[dashed] (-1.4889,1.4889) -- (1.4,-1.4);

% Distance - D_s
\node at (0.7,1.1) {$D_s$};
\draw[<->] (po1) -- ($ (po1)+(45:1cm) $); 
% Distance - D
\node at (-0.6,-0.2) {$D$};
\draw[<->] (po2) -- ($ (po2)+(145:0.6cm) $); 

\end{tikzpicture}
\caption{Infeasibility of the QP algorithm by extending the safety margin to handle the velocity-tracking error.}
\label{figure.example.infeasible}
\end{figure}
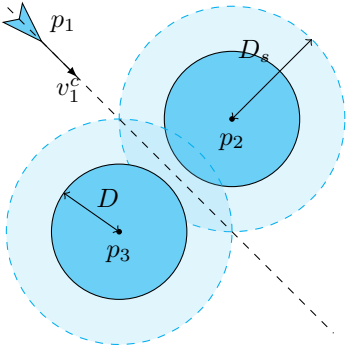

\begin{example}[Possible infeasibility of a conventional QP algorithm by extending safety margin to handle velocity-tracking error]\label{example.infeasible}
Consider the scenario involving one mobile agent with position $p_1$ and two obstacles with positions $p_2$ and $p_3$; see Figure \ref{figure.example.infeasible}. In this example, we consider $p_2=[0.5; 0.5]$, $p_3=-[0.5; 0.5]$, and $v_1^c=[1;-1]$. If the velocity-tracking error is neglectable, then one may consider
\begin{align}
%v_1^*=\argmin_{v_1^*\in\mathcal{P}_1^o}|v_1^*-v_1^c| \label{eq.example.infeasible.QP} \\
v_1^*=\argmin_{v_1^*\in\mathcal{P}_1^o} \frac{1}{2}v^{*T}_1v^{*}_1 - v^{cT}_1v^{*}_1 \label{eq.example.infeasible.QP} 
\end{align}
where $\mathcal{P}_1^o=\{v_1^*\in\mathbb{R}^n:A_1^o v_1^*+a_1^o\le 0\}$ with $A_1^o=[-\hat{\tilde{p}}_{12}, -\hat{\tilde{p}}_{13}]^T$ and $a_1^o=[|\tilde{p}_{12}|^{-1}-D^{-1}, |\tilde{p}_{13}|^{-1}-D^{-1}]^T$.

When the velocity-tracking error is nonzero, one may consider extending the safety margin and redesigning the safety controller with new $a_1^o=[|\tilde{p}_{12}|^{-1}-D^{-1}_s;|\tilde{p}_{13}|^{-1}-D^{-1}_s]$. But, such a treatment may result in an empty feasible set in specific cases. 
For example, in the case of $D=0.6$ and $D_s=1$, when $p_1=[0;0]$, the feasible set $\mathcal{P}_1^o$ in \eqref{eq.def.QP.conventional.feasibleset} should satisfy $\sqrt{2}^{-1}[1,1;-1,-1]v^*\le(\sqrt{2}-1)[1;1]$, and thus is empty.
\hfill $\Diamond$
\end{example}

\subsection{Introducing a Relaxation Parameter}
\label{subsection.addingrelaxationparameters}

An alternative solution is to introduce a relaxation parameter \cite{Ames-Xu-Grizzle-Tabuada-TAC-2017,Wang-Ames-Egerstedt-TRO-2017,Jankovic-Auto-2018} to the QP algorithm as
\begin{align}
%[v_i^*;\delta_i]=\argmin_{[v_i^*;\delta_i]\in\mathcal{P}_i^s(p)}|[v^{*}_i;\delta_i]-[v_i^c;\delta]|^2\label{eq.def.QP.non_Lipschitz}
\left[\begin{array}{@{}c@{}} v^{*}_i\\ \delta_i \end{array}\right] = \argmin_{[v_i^*;\delta_i]\in\mathcal{P}_i^s(p)}
\frac{1}{2} \left[\begin{array}{@{}c@{}} v^{*}_i\\ \delta_i \end{array}\right]^T\left[\begin{array}{@{}c@{}} v^{*}_i\\ \delta_i \end{array}\right]
-  \left[\begin{array}{@{}c@{}} v_i^c\\ \delta \end{array}\right]^T\left[\begin{array}{@{}c@{}} v^{*}_i\\ \delta_i \end{array}\right] \label{eq.def.QP.non_Lipschitz}
\end{align}
where $\delta_i$ is the relaxation parameter, 
%$\delta$ is a design parameter intended to minimize the impact of $\delta_i$,
$\delta$ is intended to compensate the effect of $\delta_i$,
\begin{align}
\mathcal{P}_i^s(p)=\left\{[v_i^*;\delta_i]:A_i^s(p)v_i^*+a_i^s(p)\delta_i\le 0,\delta_i\in[0,\delta]\right\}\label{eq.def.QP.non_Lipschitz.feasibleset}
\end{align}
is the feasible set with $A_i^s(p)=A^o_i(p)$ defined in \eqref{eq.QP2.conventional.Ai}, and without loss of generality,
\begin{align}
a_i^s(p)&=\frac{1}{\delta}a_i^o(p)\label{eq.QP2.domian.ai}
\end{align}
where $a_i^o(p)$ is defined in \eqref{eq.QP2.conventional.ai}.
% and $\delta$ is intended to compensate the effect of $\delta_i$.

Clearly, if $\tilde{p}_{ij}\neq0$ for $i\in\mathcal{N}_a$ and $j\in\mathcal{N}_{ao}\setminus\{i\}$, then zero is always an element of $\mathcal{P}_i^s(p)$ and thus the QP algorithm is always feasible. However, such a modification does not guarantee the Lipschitz continuity of the solution; see Example \ref{example.Not Lipschitz}. 
% As shown in the numerical example in Subsection \ref{subsection.experiment.1+2}, discontinuous $v_i^*$ may destroy the safety of the mobile agent, due to the uncertain actuation dynamics.

\begin{figure}[h!]
\centering
\includegraphics[width=0.8\linewidth]{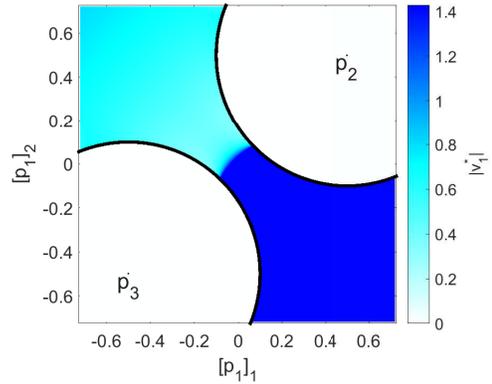}
\caption{$|v_1^*|$ with respect to different $p_1$: Lipschitz continuity of the solution to the QP algorithm cannot be guaranteed by simply adding a relaxation parameter.}
\label{fig:simulationnonlipschitz}
\end{figure}

\begin{example}[Lipschitz continuity of the solution to a conventional QP algorithm not guaranteed by simply adding a relaxation parameter]
\label{example.Not Lipschitz}
Still consider the case in Example \ref{example.infeasible}. We use $(\breve{A}_i^s,\breve{a}_i^s)$ to represent the non-redundant active constraints\footnote{A constraint of the QP problem is non-redundant if removing it changes the feasible set; it is active at a solution $v_i^*$ to the QP problem, if its equality holds at $v_i^*$ \cite{Boyd-Vandenberghe-book-2004}.} of the QP algorithm defined by \eqref{eq.def.QP.non_Lipschitz}, with $\breve{A}_i^s$ and $\breve{a}_i^s$ being submatrices of $A_i^s$ and $a_i^s$, respectively. From \cite[Example 2.1.5]{Bertsekas-book-1997}, the solution to the QP algorithm is
\begin{align}
v_1^*=v^c_1-\breve{A}_1^{sT}(\breve{A}_1^s \breve{A}_1^{sT}+\breve{a}_1^s\breve{a}_1^{sT})^{-1}(\breve{A}_1^sv_1^c+\breve{a}_1^s\delta).
\end{align}
Suppose that the non-redundant active constraints are not changed in some domain of $p$. 
Then, in this domain, taking the partial derivative of $v_1^*$ with respect to $\breve{a}_1^s$ yields
\begin{align}
\frac{\partial v_1^*}{\partial\breve{a}_1^{sT}} =& \frac{\delta\Lambda_2(2\Lambda_3\Lambda_0^{-1}-\Lambda_1)-2\Lambda_2\Lambda_3\Lambda_2^Tv_1^c\breve{a}_1^{sT}\Lambda_0^{-1}}{\Lambda_1^2}+\notag\\
&\frac{(v_1^{cT}\Lambda_2\breve{a}^{s}_1)\otimes\Lambda_2+(v_1^{cT}\Lambda_2) \otimes(\Lambda_2\breve{a}_1^s)}{\Lambda_1}\label{eq.expamle.partialMa}
\end{align}
where $\Lambda_0=\breve{A}_1^{s}\breve{A}_1^{sT}$, $\Lambda_1=(1+\breve{a}^{sT}_1\Lambda_0^{-1}\breve{a}^{s}_1)$, $\Lambda_2= \breve{A}^{sT}_1\Lambda_0^{-1}$ and $\Lambda_3=\breve{a}^{s}_1\breve{a}^{sT}_1$. 
%As a special case, we consider $v^c_1=0$ and $\breve{a}^{s}_1 = 0$. 
%$\breve{A}^{s}_1$ is some rows in $A^s_1$, which is defined in \eqref{eq.example.infeasible.QP} and consisted of the relative positions between agent and obstacles.
$\breve{A}^{s}_1$ is consisted of the relative positions between the agent and the obstacles.
Thus, $\sigma_{\min}(\breve{A}^{s}_1)$ cannot be guaranteed to be lower bounded by a positive constant. This means that the solution to the QP algorithm may not be Lipschitz. Indeed, \cite[Theorem 3.1]{Hager-SIAMControl-1979} requires a positive lower bound of $\sigma_{\min}(\breve{A}^{s}_1)$ to guarantee the Lipschitz continuity of the QP problem. In this numerical example, we consider $D = 0.6$, $D_s = 0.68$, $v^c_1 = [1;-1]$, $p_1 \in \{p:|p-p_{2}|\ge D, |p-p_{3}|\ge D~\}$, $p_2=[0.5; 0.5]$, $p_3=-[0.5; 0.5]$,
$\delta = 100$, $A^s_1(p)=[-\hat{\tilde{p}}_{12}, -\hat{\tilde{p}}_{13}]^T$ and $a^s_1(p) = \delta^{-1}\left[|\tilde{p}_{12}|^{-1}- D^{-1}_s, |\tilde{p}_{12}|^{-1}- D^{-1}_s\right]^T$. Figure \ref{fig:simulationnonlipschitz} shows how $p_1$ influences $|v_1^*|$. One may observe the sudden change of $|v_1^*|$ when $p_1$ is close to zero.
\hfill $\Diamond$
\end{example}

Examples \ref{example.infeasible} and \ref{example.Not Lipschitz} show that the methods mentioned above do not guarantee the local Lipschitz property of the solution to the QP algorithm with respect to time.
However, the local Lipschitz continuity of the solution is viewed to be essential for the robustness of a QP algorithm in practical systems.
In the system setup in this paper, the non-Lipschitz velocity reference signal may lead to a large gain from the barrier function to the velocity-tracking error and result in the unexpected dynamic response of the actuation system, which may violate the safety of the nominal system. 
This is also verified by the numerical example in Subsection \ref{subsection.experiment.1+2}.

%{\color{red} \bf
%Under the non-Lipschitz velocity reference, it is difficult to guarantee the robustness of safety with respect to uncertain actuation dynamics. 
%And the non-Lipschitz velocity reference signal generated by the QP algorithm may leads to an unexpected response and causes collision.
%Indeed, some related studies also note the velocity reference should admit locally Lipschitz \cite{Wang-Ames-Egerstedt-ICRA-2017, Xu-Tabuada-Grizzle-Ames-IAC-2015}.
%The proposed method is given in Section \ref{section.reshapefeasibleset}.
%%However, Assumption \ref{assumption.multi.referencetracking} requires the Lipschitz property.
%%If the solution is not locally Lipschitz continuous, the small-gain theorem cannot be applied to analyze the boundedness of velocity-tracking error.
%%The proposed method is given in Section \ref{section.reshapefeasibleset}.
%}

\section{New QP-based Design with Relaxation Parameter and Reshaped Feasible Region}
\label{section.reshapefeasibleset}

In this section, we present our solution to the safety control problem involving multiple controlled mobile agents and multiple stationary obstacles.
Our major contribution lies in a new class of QP-based controllers with relaxation parameters and reshaped feasible sets to address the non-Lipschitz issue discussed in Section \ref{section.motivatingexamples}. 
To be specific, positive bases\footnote{
A positive combination of the set of vectors $\{a_j\in\mathbb{R}^n:j=1,\ldots,r\}$ is a linear combination $\lambda_1a_1+\cdots+\lambda_ra_r$ with $\lambda_j\ge 0$; it is a strictly positive combination if $\lambda_j>0$ for $j=1,\ldots, r$. 
A set of vectors $\{a_1,\ldots a_r\}$ is positively dependent if one of the $a_i$ is a positive combination of the others; otherwise, the set is positively independent. 
A positive basis for a subspace $\mathcal{C}\subseteq\mathbb{R}^n$ is a set of positively independent vectors whose span is $\mathcal{C}$\cite{Davis-AJM-1954}.
} are used to reshape the feasible sets for ensured feasibility and Lipschitz continuity.

The proposed safety controller is in the form of
\begin{align}
%[v^{*}_i;\delta_i]&=\argmin_{[v^{*}_i;\delta_i]\in\mathcal{P}^{r}_i(p)} |[v^{*}_i; \delta_i]-[v^c_i;\delta]|^2,\label{eq.def.calP'_i}
\left[\begin{array}{@{}c@{}} v^{*}_i\\ \delta_i \end{array}\right] = \argmin_{[v_i^*;\delta_i]\in\mathcal{P}_i^r(p)}
\frac{1}{2} \left[\begin{array}{@{}c@{}} v^{*}_i\\ \delta_i \end{array}\right]^T\left[\begin{array}{@{}c@{}} v^{*}_i\\ \delta_i \end{array}\right]
-  \left[\begin{array}{@{}c@{}} v_i^c\\ \delta \end{array}\right]^T\left[\begin{array}{@{}c@{}} v^{*}_i\\ \delta_i \end{array}\right] \label{eq.def.calP'_i}
\end{align}
where 
\begin{align}
\mathcal{P}^{r}_i(p)=\left\{[v^{*}_i;\delta_i]:~A^r_iv^{*}_i + a^r_i(p)\delta_i\le 0,\delta_i\in[0, \delta]\right\}\label{eq.def.calP'_i.feasibleset}
\end{align}
is the reshaped feasible set. Here, $A^r_i\in\mathbb{R}^{n_p\times n}$ is a constant matrix with $n_p>n$, and satisfies that each row is a unit vector, any $n$ rows are linearly independent, and for any unit vector $\hat{u}\in\mathbb{R}^n$,
\begin{align}
\min_{q\in\mathcal{Q}(\hat{u})}\max_{j\in\mathcal{J}(\hat{u})} [A^r_i]_{j,:} q &> 0, \label{eq.condition.Api}\\
\card(\mathcal{J}(\hat{u})) &\ge n, \label{eq.condition.J}
\end{align}
where $\mathcal{Q}(\hat{u}) = \{q\in\mathbb{R}^n : q^T\hat{u} > 0 \}$, $\mathcal{J}(\hat{u}) = \{j=1,\ldots,n_p : [A^r_i]_{j,:} \hat{u} \ge c_A \}$ with $c_A$ being a positive constant less than $1$, and $\card$ takes the cardinality.

We choose $a^r_i:\mathbb{R}^{n\times n_{ao}}\rightarrow\mathbb{R}^{n_p}$ as
\begin{align}
a^r_i(p)&=\max\left\{\varphi\left(A^s_i(p),a^s_i(p),\frac{c_P}{\delta}\right)\right\}\label{eq.reshapedfeasibleset.api}
\end{align}
with 
\begin{align}
&\varphi\left(A^s_i,a_i^s,\frac{c_P}{\delta}\right)=A^r_iA^{sT}_i\odot a_i^{sT}\notag\\
&~~~-\sat(c_K(c_A-A^r_iA^{sT}_i))\odot(A^r_iA^{sT}_i\odot a_i^{sT}+\frac{c_P}{\delta}), \label{eq.chi.A_ia_i} 
\end{align}
where $c_A$ is associated with $A^r_i$ given in the definition of $\mathcal{J}(\hat{u})$, {\color{blue}$c_K\ge c_A^{-1}$} and $c_P$ can be any positive constants, and $\delta$, $A^s_i$ and $a^s_i$ are defined in  \eqref{eq.def.QP.non_Lipschitz}--\eqref{eq.def.QP.non_Lipschitz.feasibleset}.

Condition \eqref{eq.condition.Api} guarantees that any vector in $\mathbb{R}^n$ can be represented by a strictly positive combination of some row vectors of $A^r_i$ \cite[Theorem 3.3]{Davis-AJM-1954}. Based on \cite[Theorem 3.1]{Hager-SIAMControl-1979}, it can be proved that such an $A^r_i$ can be used to guarantee Lipschitz continuity of the solution to the QP algorithm. The existence of such an $A^r_i$ is proved in Subsection \ref{subsection.Aipexistence} as part of the proof of the main result in Theorem \ref{theorem.smallgaincondition2}.

In Example \ref{example.feasible_region}, we examine the refined QP algorithm \eqref{eq.def.calP'_i} in the same scenario as in Examples \ref     {example.infeasible} and \ref{example.Not Lipschitz}.

\begin{figure}[h!]
\centering
% Do not use scale for this tikzpicture. 
% Do not use scale for this tikzpicture. 
\includegraphics[width=0.8\linewidth]{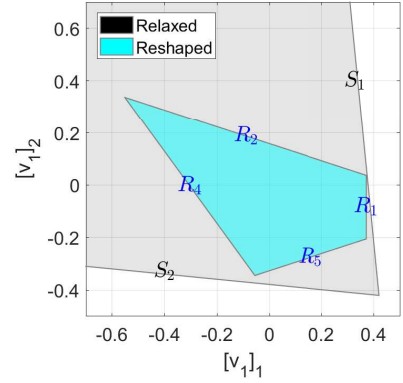}
\caption{The relaxed and the reshaped feasible sets with $S_i:~[A^s_1]_{i,:}v^*_1 = -[a^s_1]_{i}\delta_1$ for $i=1,2$ and $R_i:~[A^r_1]_{i,:}v^*_1 = -[a^r_1]_{i}\delta_1$ for $i=1,\ldots,5$. $R_3$ corresponds to a redundant constraint, and is outside the range of the figure.}
\label{figure.reshape_description}
\end{figure}

\begin{figure}[h!]
\centering
\includegraphics[width=0.8\linewidth]{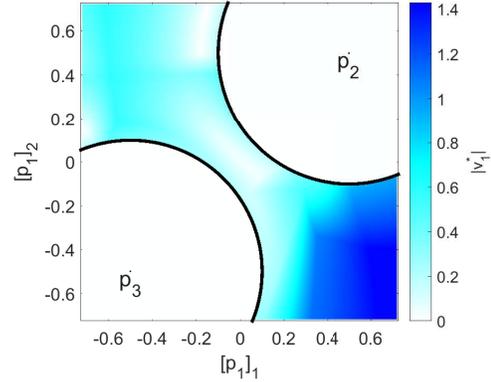}
\caption{$|v_1^*|$ with respect to different $p_1$: In the scenario of Example \ref{example.Not Lipschitz}, the Lipschitz continuity of the solution to the QP algorithm can be guaranteed by adding a relaxation parameter and reshaping the feasible set.}
\label{fig:simulationlipschitz}
\end{figure}

\begin{example}[Lipschitz continuity of the solution to the QP algorithm guaranteed by adding a relaxation parameter and reshaping the feasible set]
\label{example.feasible_region}
Continue Example \ref{example.Not Lipschitz}. We construct a $\mathcal{P}^{r}_i$ in the form of \eqref{eq.def.calP'_i.feasibleset} and examine the solution to the QP algorithm \eqref{eq.def.calP'_i} with a reshaped feasible set. For any specific odd integer $n_p \ge 5$, a typical $A^r_i\in\mathbb{R}^{n_p \times 2}$ consists of the outward normal vectors to an odd-sided regular polygon, that is, each row of $A^r_i$ is $[\cos(2\pi j/n_p),~\sin(2\pi j/n_p) ]$ for $j=1,\ldots,n_p$. Accordingly, $c_A = \cos(2\pi/n_p)$.
In the numerical simulation, we choose $n_p = 5$, $c_K=1$, $c_P = 5/2$ and $c_A = \cos(2\pi/5)$. Figure \ref{figure.reshape_description} shows the original and the reshaped feasible sets when $p_1=[-0.4;0.4]$. Figure \ref{fig:simulationlipschitz} shows how $p_1$ influences $|v_1^*|$, which is in accordance with our expectation of Lipschitz continuity.\hfill $\Diamond$
\end{example}

When the relaxation parameter $\delta_i$ goes to zero, the robustness with respect to the velocity-tracking error $\tilde{v}_i$ is weakened (see equation \eqref{eq.Vij.derivative} in the proof of Proposition \ref{proposition.multi.robustISpSlyapunov}). In this case, motivated by \cite{Wang-Ames-Egerstedt-TRO-2017}, we set $v_i^*=0$. 
Hence, the implemented safety controller is defined as
\begin{align}
\hspace{-5pt}\left[\begin{array}{@{}c@{}}v^*_i(t) \\ \delta_{i}(t) \end{array}\right]\!=\!\begin{cases}
%\argmin_{[v^{*}_i; \delta_i] \in \mathcal{P}^{r}_i(p)}|[v^{*}_i; \delta_i] - [v^c_i; \delta]|^2,\\
\argmin\limits_{[v^{*}_i; \delta_i] \in \mathcal{P}^{r}_i(p)} \dfrac{1}{2} \left[\begin{array}{@{}c@{}} v^{*}_i\\ \delta_i \end{array}\right]^T\!\left[\begin{array}{@{}c@{}} v^{*}_i\\ \delta_i \end{array}\right]\! -\! \left[\begin{array}{@{}c@{}} v_i^c\\ \delta \end{array}\right]^T\!\left[\begin{array}{@{}c@{}} v^{*}_i\\ \delta_i \end{array}\right], \\
~~~~~~~~~~~~~~~~~~~~~~\text{if~}\inf_{\tau\in[0,t)}\delta_i(\tau)\neq 0\\
0,~~~~~~~~~~~~~~~~~~~\text{if~}\inf_{\tau\in[0,t)}\delta_i(\tau)= 0 
\end{cases}\hspace{-5pt}\label{eq.def.v*_i.controller}
\end{align}
where $\mathcal{P}^{r}_i$ is defined in \eqref{eq.def.calP'_i}.

For convenience of discussions later, we denote
\begin{align}
T_i=\min\{t\in\mathbb{R}_+:\delta_i(t)=0\}\label{eq.def.Ts_i}
\end{align}
as the braking time of mobile agent $i$, and set $T_i=0$ for $i\in\mathcal{N}_o$.
We use
\begin{align}
\mathcal{N}_{s}(t) = \{i=1,\ldots, n_{a}:\delta_i(t)=0\}
\end{align}
to represent the set of the breaking agents, and denote
\begin{align}
V_m&=\max_{i\in\mathcal{N}_a,j\in\mathcal{N}_{ao}\setminus \{i\}}V_{ij},\label{eq.Vm.definition}\\
V_R&=\max_{i\in\mathcal{N}_a\setminus\mathcal{N}_s,j\in\mathcal{N}_{ao}\setminus \{i\}}V_{ij},\label{eq.V.running.definition}\\
V_S&=\max_{i\in\mathcal{N}_s,j\in\mathcal{N}_s\cup\mathcal{N}_o\setminus \{i\}}V_{ij}, \label{eq.V.breaking.definition}\\
\tilde{v}_m&=\argmax_{x\in\{\tilde{v}_i:i\in\mathcal{N}_a\}}|x|,\label{eq.def.tildevm}\\
z_m&=\argmax_{x\in\{z_i:i\in\mathcal{N}_a\}}|x|.\label{eq.zm.definition}
\end{align}

Our main result is given by Theorem \ref{theorem.smallgaincondition2}. 
% It is shown that one can design collision avoidance controllers in the form of \eqref{eq.def.v*_i.controller} such that collision avoidance is guaranteed by the controlled mobile agents as long as specific initial conditions are satisfied.

\begin{theorem}\label{theorem.smallgaincondition2}
Under Assumptions \ref{assumption.multi.referencetracking} and \ref{assumption.multi.boundedprimarycontrol}, consider the multi-agent system \eqref{eq.def.p_i.kinematics}--\eqref{eq.def.v_i.actuation} and the safety controller \eqref{eq.def.calP'_i}--\eqref{eq.def.v*_i.controller}. There exist $\mu\in\mathcal{M^-C}$, $\alpha_V\in\mathcal{K}^e$ and positive constants $\delta$, $D$, $D_s$, $\bar{v}^c$, $\bar{v}^{cd}$, $\bar{z}_{m0}$ and $\bar{V}_{m0}$, such that for any initial state satisfying $|z_m(0)|\leq\bar{z}_{m0}$ and $V_m(0)\leq\bar{V}_{m0}$, property \eqref{eq.problemformulation} holds for all $t\geq 0$.
\end{theorem}

The proof of Theorem \ref{theorem.smallgaincondition2} is given in Section \ref{section.proofs}.

\begin{remark}
For the special case involving a single agent and a single obstacle as shown in Figure \ref{figure.collisionfree}, it can be directly verified that the conventional QP algorithm \eqref{eq.def.QP.conventional} satisfies all the conditions given by \cite[Theorem 3.1]{Hager-SIAMControl-1979}, and thus, the solution is Lipschitz. In this case, adding relaxation parameters and reshaping the feasible set are unnecessary. Still, the idea of considering the controlled multi-agent system as an interconnected system in Section \ref{section.reshapefeasibleset} is still valid even if the obstacle is moving. A detailed discussion can be found in \cite{technicalreport}.
\end{remark}

\section{Properties of the Proposed Design and Proof of the Main Result}
\label{section.proofs}

This section proves Theorem \ref{theorem.smallgaincondition2} by observing new properties of the proposed refined QP-based controller. We first show the existence of $A^r_i$ for feasible-set reshaping (Subsection \ref{subsection.Aipexistence}) and prove that the reshaped feasible set is a subset of the feasible set with relaxation parameters (Subsection \ref{subsection.reshape-subset}). Then, we use gains to describe the interconnections between the controlled nominal systems and the uncertain actuation systems (Subsections \ref{subsection.multi.QP} and \ref{subsection.actuationresponse}) and present a small-gain analysis to guarantee the safety of the controlled multi-agent system (Subsection \ref{subsection.smallgain}).

\subsection{Existence of $A^r_i$ for Feasible-Region Reshaping}
\label{subsection.Aipexistence}

\begin{proposition}\label{proposition.Aipexistence}
There exist an $A^r_i\in\mathbb{R}^{n_p\times n}$ with $n_p> n$, any $n$ rows of which are linearly independent, such that for any unit vector $\hat{u}\in\mathbb{R}^n$, condition \eqref{eq.condition.Api}--\eqref{eq.condition.J} is satisfied.
\end{proposition}

Due to space limitation, the proof of Proposition \ref{proposition.Aipexistence} is given in the technical report \cite{technicalreport}.

\subsection{Reshaped Feasible Region Belonging to the Original Feasible Region}
\label{subsection.reshape-subset}

The following proposition shows that the reshaped feasible set $\mathcal{P}^r_i$ is a subset of the relaxed feasible set $\mathcal{P}_i^s$.

\begin{proposition}\label{proposition.calP_i.subset}
Consider $\mathcal{P}_i^s(p)$ defined by \eqref{eq.def.QP.non_Lipschitz} and $\mathcal{P}^{r}_i(p)$ defined by \eqref{eq.def.calP'_i}. We have $\mathcal{P}^{r}_i(p)\subseteq\mathcal{P}_i^s(p)$.
\end{proposition}

\begin{IEEEproof}
With $v^{*}_i\in\mathbb{R}^n$ and $\delta_i\in[0, \delta]$, we define
\begin{align}
\mathcal{H}^s_{ik}(p)&=\left\{[v^{*}_i;\delta_i]:[A^s_i(p)]_{k,:}v^{*}_i+[a^s_i(p)]_{k,:}\delta_i \le 0\right\}\label{eq.def.calH1}\\
\mathcal{H}^{r}_{ik}(p)&=\left\{[v^{*}_i;\delta_i]:\right. \notag\\ 
&\left. A^r_i v^{*}_i+\varphi([A^s_i(p)]_{k,:}, [a^s_i(p)]_{k,:}, c_P/\delta)\delta_i \le 0 \right\}\label{eq.def.calH2}
\end{align}
for $k=1,\ldots,n_{ao}-1$, where $\varphi$ is defined in \eqref{eq.chi.A_ia_i}.

Obviously, $\mathcal{H}^s_{ik}(p)$ corresponds to the $k$-th constraint of the feasible set $\mathcal{P}^s_i(p)$, and thus
\begin{align}
\mathcal{P}^s_i(p)=\bigcap^{n_{ao}-1}_{k=1}\mathcal{H}^s_{ik}(p).\label{eq.def.Pi}
\end{align}
%And, $\mathcal{H}^{r}_{ik}(p)$ reshaped the corresponds to the $k$-th constraint of the feasible set $\mathcal{P}^s_i(p)$

The definition of $\mathcal{P}^r_i$ in \eqref{eq.def.calP'_i.feasibleset} implies that
\begin{align}
%A^r_iv^{*}_i+\varphi_i^{k}(p)\delta_i\le \varphi_i^{k}(p)\delta_i-a^r_i(p)\delta_i.\label{eq.def.temp}
A^r_iv^{*}_i+a^r_i(p)\delta_i \le 0 \label{eq.def.temp}
\end{align}
holds for any $[v^*_i;\delta_i]$ in $\mathcal{P}^r_i(p)$. 
We rewrite $\varphi$ defined in \eqref{eq.chi.A_ia_i} as
\begin{align}
&\varphi(A^s_i(p),a^s_i(p),c_P/\delta)%=[\varphi_i^1(p),\ldots,\varphi_i^{n_p}(p)],
=\notag \\
&\left[\begin{array}{l}
\varphi^T([A^s_i(p)]_{1,:}, [a^s_i(p)]_{1,:}, c_P/\delta)\\
~~~~~~~~~~~~~~~~\vdots\\
\varphi^T([A^s_i(p)]_{n_{ao}-1,:}, [a^s_i(p)]_{n_{ao}-1,:}, c_P/\delta)
\end{array}\right]^T,
\end{align}
which together with the definition of $a^r_i$ in \eqref{eq.reshapedfeasibleset.api} implies that
\begin{align}
\varphi([A^s_i(p)]_{k,:}, [a^s_i(p)]_{k,:}, c_P/\delta) \le a^r_i(p).\label{eq.proof.chi_ik}
\end{align}
for all $k\in\{1,\ldots,n_{ao}-1\}$.
Then, by using $\delta_i\in[0,\delta]$ and combining \eqref{eq.def.temp} and \eqref{eq.proof.chi_ik}, we have $\mathcal{P}^r_i\subseteq\mathcal{H}^r_{ik}(p)$ and thus
\begin{align}
\mathcal{P}^r_i(p)\subseteq\bigcap^{n_{ao}-1}_{k=1}\mathcal{H}^r_{ik}(p).\label{eq.def.Pi'}
\end{align}

Based on \eqref{eq.def.Pi} and \eqref{eq.def.Pi'}, we have
\begin{align}
&\mathcal{H}^r_{ik}(p)\subseteq\mathcal{H}^s_{ik}(p)~\text{for}~k=1,\ldots,n_{ao}-1\nonumber\\
\Rightarrow &\mathcal{P}^{r}_i(p)\subseteq\mathcal{P}^s_i(p).\label{eq.proof.def.temp5}
\end{align}

Now we prove $\mathcal{H}^r_{ik}(p)\subseteq\mathcal{H}^s_{ik}(p)$.

For $k=1,\ldots,n_{ao}-1$, denote
\begin{align}
\mathcal{V}([A^s_i(p)]_{k,:})&=\{[A^r_i]_{q,:}:~q\in\mathcal{J}([A^s_i(p)]_{k,:})\}.
\end{align}
Then, from the conditions for the definition of $A^r_i$ given by \eqref{eq.condition.Api} and \eqref{eq.condition.J}, $[A^s_i(p)]_{k,:}$ is a positive combination of some $n$ elements of $\mathcal{V}([A^s_i(p)]_{k,:})$ \cite[Theorem 3.3]{Davis-AJM-1954}. We use these $n$ elements to form the rows of matrix $C^{r}_{ik}$, and define
\begin{align}
\bar{\mathcal{H}}^{r}_{ik}(p)=\{[v^*_i;\delta_i]:C^{r}_{ik} v^*_i+C^{r}_{ik}[A^s_i(p)]^T_{k,:}[a^s_i(p)]_{k,:}\delta_i\le 0\}.\label{eq.proof.def.S}
\end{align}
By using the definitions of $\varphi$ in \eqref{eq.chi.A_ia_i} and $\mathcal{H}^r_{ik}(p)$ in \eqref{eq.def.calH2}, we have
\begin{align}
\mathcal{H}^r_{ik}(p)\subseteq\bar{\mathcal{H}}^{r}_{ik}(p).\label{eq.def.temp3}
\end{align}

Since any $n$ rows of $A^r_i$ are linearly independent, by using the properties of positive combinations, we have $[A^s_i(p)]_{k,:}C^{r^{-1}}_{ik}\ge 0$. Multiplying both sides of the constraint inequality of $\bar{\mathcal{H}}^{r}_{ik}(p)$ in \eqref{eq.proof.def.S} by $[A^s_i(p)]_{k,:}C^{r^{-1}}_{ik}$ yields
\begin{align}
&[A^s_i(p)]_{k,:}C^{r^{-1}}_{ik}\left(C^{r}_{ik} v^*_i+C^{r}_{ik}[A^s_i(p)]^T_{k,:}[a^s_i(p)]_{k,:}\delta_i\right)\nonumber\\
&=[A^s_i(p)]_{k,:}v^{*}_i+[a^s_i(p)]_{k,:}\delta_i \le 0.
\end{align}
Clearly, the last inequality is in accordance with the constraint inequality of $\mathcal{H}^s_{ik}(p)$ defined in \eqref{eq.def.calH1}, which means
\begin{align}
\bar{\mathcal{H}}^{r}_{ik}(p)\subseteq\mathcal{H}^s_{ik}(p). \label{eq.def.temp4}
\end{align}
Properties \eqref{eq.def.temp3} and \eqref{eq.def.temp4} together imply
\begin{align}
%\mathcal{H}^s_{ik}(p)\subseteq\mathcal{H}^k_i(p).
\mathcal{H}^r_{ik}(p)\subseteq\mathcal{H}^s_{ik}(p).
\end{align}
Recall \eqref{eq.proof.def.temp5}. This completes the proof of Proposition \ref{proposition.calP_i.subset}.
\end{IEEEproof}

\subsection{Robustness of the Nominal System Safety}
\label{subsection.multi.QP}

The following proposition gives the range of the relaxation parameter $\delta_i$, which is to be used for the robustness analysis later.

\begin{proposition}\label{proposition.relaxation_parameter}
For any $\tilde{p}_{ij} \neq 0$ with $i\in\mathcal{N}_a$ and $j\in\mathcal{N}_{ao}\setminus\{i\}$, the solution $[v^*_i;\delta_i]$ to the QP algorithm \eqref{eq.def.calP'_i}--\eqref{eq.chi.A_ia_i} has the following properties:
\begin{enumerate}
\item If $\delta_i=0$, then $v^*_i=0$;
\item If $\delta_i>0$, then $a^r_i(p)$ is bounded;
\item By choosing $\delta$ large enough, it can be guaranteed that $\delta_i>0\Rightarrow\delta_i\ge \delta/2$.
\end{enumerate}
\end{proposition}

\begin{IEEEproof}
The properties are proved one-by-one.

{\bfseries Property 1:} For any specific $\delta_i\ge 0$ and any $\tilde{p}_{ij} \neq 0$ with $i\in\mathcal{N}_a$ and $j\in\mathcal{N}_{ao}\setminus\{i\}$, define
\begin{align}
\bar{\mathcal{P}}^{r}_i(p)=\left\{v^{*}_i:A^r_iv^{*}_i \le-a^r_i(p)\delta_i\right\}.\label{v^*.st}
\end{align}
Then, the definition of $A^r_i$ below \eqref{eq.def.calP'_i.feasibleset} guarantees that $\bar{\mathcal{P}}^{r}_i$ is a bounded, closed, convex polyhedron\footnote{A polyhedron is the intersection of a finite number of half-spaces and hyperplanes \cite{Boyd-Vandenberghe-book-2004}.}.

Suppose that there exists some $d\neq 0$ such that $A^r_id\leq0$. Then, for any $v_i^*\in\bar{\mathcal{P}}^{r}_i(p)$ and any $\epsilon>0$,
\begin{align}
A^r_i(v_i^*+\epsilon d)=A^r_iv_i^*+\epsilon A^r_id\le -a^r_i(p)\delta_i,
\end{align}
i.e., $(v_i^*+\epsilon d)\in \bar{\mathcal{P}}^{r}_i(p)$. This means that $\bar{\mathcal{P}}^{r}_i(p)$ is not bounded.

By contradiction, $v_i^*$ is the only solution of the constraint $A^r_iv_i^*\leq 0$. Property 1 is proved.

{\bfseries Property 2:} Recall the definitions of $\mathcal{P}^s_i$, $\mathcal{P}^r_i$ and $\mathcal{H}^s_{ik}$ in \eqref{eq.def.QP.non_Lipschitz.feasibleset}, \eqref{eq.def.calP'_i.feasibleset} and \eqref{eq.def.calH1}, respectively. From the proof of Proposition 2, we have
\begin{align}
\mathcal{P}^r_i(p)\subseteq\mathcal{P}^s_i(p)\subseteq\mathcal{H}^s_{ik}(p).\label{eq.Aipkv*.subsets}
\end{align}

For $k=1,\ldots, n_{ao}-1$, denote
\begin{align}
\mathcal{V}(-[A^s_i(p)]_{k,:})=\left\{[A^r_i]_{q,:}:~ q\in \mathcal{J}(-[A^s_i(p)]_{k,:})\right\}.
\end{align}
Then, from the properties of $A^r_i$ given by \eqref{eq.condition.Api} and  \eqref{eq.condition.J}, $-[A^s_i(p)]_{k,:}$ is a positive combination of some $n$ elements of $\mathcal{V}(-[A^s_i(p)]_{k,:})$. We define matrix $C^{r}_{ik}$ with these $n$ elements as the rows, define vector $a^r_{ik}$ with the elements of $a^r_i(p)$ corresponding to these $n$ rows, and define
\begin{align}
\tilde{\mathcal{H}}^{r}_{ik}(p)=\{[v^*_i,\delta_i]:C^{r}_{ik} v^*_i+a^r_{ik}(p)\delta_i\le 0\}.\label{barH}
\end{align}
By using the definition of $\mathcal{P}^{r}_i$ in \eqref{eq.def.calP'_i.feasibleset}, we have
\begin{align}
\mathcal{P}^{r}_i(p)\subseteq \tilde{\mathcal{H}}^{r}_{ik}(p).\label{eq.Aipkv*.subsets2}
\end{align}

Combining \eqref{eq.Aipkv*.subsets} and \eqref{eq.Aipkv*.subsets2} implies
\begin{align}
\mathcal{P}^{r}_i(p)\subseteq\mathcal{H}^s_{ik}(p)\cap\tilde{\mathcal{H}}^{r}_{ik}(p).
\end{align}

Since any $n$ rows of $A^r_i$ are linearly independent, by using the property of positive combinations, we have 
\begin{align}
-[A^s_i(p)]_{k,:}C^{r^{-1}}_{ik}=:k_s\ge 0. \label{eq.def.k_s}
\end{align}
Multiplying both sides of the constraint inequality of $\tilde{\mathcal{H}}^{r}_{ik}(p)$ in \eqref{barH} by the nonnegative $k_s$ defined in \eqref{eq.def.k_s} yields
\begin{align}
-[A^s_i(p)]_{k,:}C^{r^{-1}}_{ik}C^r_{ik} v^*_i+k_sa^r_{ik}\delta_i&=-[A^s_i(p)]_{k,:} v^*_i+k_sa^r_{ik}\delta_i\nonumber\\
&\le 0.\label{eq.Aipkv*.range2}
\end{align}
Also, the definition of $\mathcal{H}^s_{ik}(p)$ in \eqref{eq.def.calH2} implies that
\begin{align}
[A^s_i(p)]_{k,:}v^*_i\le -[a^s_i(p)]_{k, :}\delta_i.\label{eq.Aipkv*.range3}
\end{align}

By using \eqref{eq.Aipkv*.range2} and \eqref{eq.Aipkv*.range3}, we have that any element $[v^*_i,\delta_i]$ of $\mathcal{H}^s_{ik}(p)\cap\tilde{\mathcal{H}}^{r}_{ik}(p)$ satisfies
\begin{align}
k_sa^r_{ik}\delta_i\le[A^s_i(p)]_{k,:}v^*_i\le-[a^s_i(p)]_{k, :}\delta_i.
\end{align}
If $k_sa^r_{ik}>-[a^s_i(p)]_{k,:}$, then the only element of $\mathcal{H}^s_{ik}(p) \cap \tilde{\mathcal{H}}^{r}_{ik}(p)$ is $0$, and thus, $\mathcal{P}^{r}_i(p)=\{0\}$. This contradicts with $\delta_i>0$. Thus, we have
\begin{align}
[a^s_i(p)]_{k,:}\leq-k_sa^r_{ik}.\label{eq.bark_s}
\end{align}
By using the definition of $\varphi$ in \eqref{eq.chi.A_ia_i}, we rewrite
%\begin{align}
%[\varphi(A_i^s,a^s_i,c_P/\delta)]_{j,k}=&c^s_{jk}[a^s_i]_{k,:}(1-\sat(c_K(c_A-c^s_{jk})))\notag \\
%&-\frac{c_P}{\delta}\sat(c_K(c_A-c^s_{jk})) \label{eq.barkbara.chijk}
%\end{align}
%where $c^s_{jk}=[A^r_iA_i^{sT}]_{j,k}$, and then we have 
%\begin{align}
%[\varphi(A_i^s,a^s_i,c_P/\delta)]_{j,k}&\ge\min\{c^s_{jk}[a^s_i]_{k,:},-\frac{c_P}{\delta}\}\notag \\&
%\ge \frac{1}{\delta}\min\{\alpha_V(-\mu(0)),-c_P\}.\label{eq.chi_jk}
%\end{align}
\begin{align}
    &[\varphi(A_i^s,a^s_i,c_P/\delta)]_{j,k}=-\frac{c_P}{\delta}\sat(c_K(c_A-[A^r_iA_i^{sT}]_{j,k}))\notag \\
    &~~~~~[A^r_iA_i^{sT}]_{j,k}[a^s_i]_{k,:}(1-\sat(c_K(c_A-[A^r_iA_i^{sT}]_{j,k}))) \label{eq.barkbara.chijk}
\end{align}
and then we have 
\begin{align}
    [\varphi(A_i^s,a^s_i,c_P/\delta)]_{j,k}&\ge\min\{[A^r_iA_i^{sT}]_{j,k}[a^s_i]_{k,:},-\frac{c_P}{\delta}\}\notag \\&
    \ge \frac{1}{\delta}\min\{\alpha_V(-\mu(0)),-c_P\}.\label{eq.chi_jk}
\end{align}
This, together with \eqref{eq.bark_s} implies
\begin{align}
[a^s_i(p)]_{k,:}&\le-k_sa^r_{ik}\le -k_s1_{n\times 1}\min \{a^r_{ik}\},\notag\\
&\le\frac{1}{\delta}\sqrt{n}|k_s|\max\{-\alpha_V(-\mu(0)), c_P\}\label{eq.barkbara.lowerbound}
\end{align}
where $1_{n\times 1}$ is the $n$-dimensional column vector of all ones.

On the other hand, by using the definitions of $a_i^o(p)$ in \eqref{eq.QP2.conventional.ai} and $a^s_i(p)$ in \eqref{eq.QP2.domian.ai}, we have
\begin{align}
[a^s_i(p)]_{k,:}=\frac{1}{\delta}\alpha_V(V_{ik}-\mu(0))\ge \frac{1}{\delta}\alpha_V(-\mu(0)).\label{eq.bound2}
\end{align}

Combining \eqref{eq.barkbara.lowerbound} and \eqref{eq.bound2} implies
\begin{align}
\frac{1}{\delta}\alpha_V(-\mu(0))&\le[a^s_i(p)]_{k, :}\nonumber\\
&\le\frac{1}{\delta}|k_s| \sqrt{n}\max\{-\alpha_V(-\mu(0)), c_P\}. \label{eq.a^s_i.bounded}
\end{align}
The boundedness of $[a^s_i(p)]_{k,:}$ for all $k=1,\ldots,n_{ao}-1$ guarantees the boundedness of $a^r_i(p)$. This completes the proof of property 2.

{\bfseries Property 3:} If $[v^{c}_i; \delta]\in \mathcal{P}^{r}_i(p)$, then property 3 is obvious. Now, we consider the case of $[v^{c}_i;\delta]\notin\mathcal{P}^{r}_i(p)$. We use $M_i=[\breve{A}^r_i,\breve{a}^r_i]$ to represent the non-redundant active constraints of the QP algorithm, with $\breve{A}^r_i$ and $\breve{a}^r_i$ being submatrices of $A^r_i$ and $a^r_i$, respectively. From \cite[Example 2.1.5]{Bertsekas-book-1997}, the solution to the QP algorithm is
\begin{align}
\left[\begin{array}{c} v^*_i \\ \delta_i \end{array}\right]=\left(I-M_i^T(M_iM_i^T)^{-1}M_i\right) \left[\begin{array}{c} v^c_i \\ \delta\end{array}\right].
\end{align}
Then, by using $M_i=[\breve{A}^r_i,\breve{a}^r_i]$, we have
\begin{align}
\delta_i
&=\delta-\breve{a}^{rT}_i(\breve{A}^r_i\breve{A}^{rT}_i+\breve{a}^r_i\breve{a}^{rT}_i)^{-1}\left(\breve{a}^r_i\delta+\breve{A}^r_i v^c_i\right)\nonumber\\
&=\frac{\delta-\breve{a}^{rT}_i\left(\breve{A}^r_i\breve{A}^{rT}_i\right)^{-1}\breve{A}^r_iv^c_i}{1+\breve{a}^{rT}_i(\breve{A}^r_i\breve{A}^{rT}_i)^{-1}\breve{a}^r_i},
\end{align}
where Sherman-Morrison-Woodbury formula \cite{Horn-Johnson-book-2012} is used for the second equality. Since $\breve{a}^r_i$ is a submatrix of $a^r_i$, according to property 2 of Proposition \ref{proposition.relaxation_parameter}, there exists a positive constant $\tau_a$ such that 
\begin{align}
|\breve{a}^r_i|\le \frac{\tau_a}{\delta}.
\end{align}

Since any $n$ rows of $A^r_i$ are chosen to be linearly independent as required after \eqref{eq.def.calP'_i.feasibleset}, there exists a positive constant $c_M$ such that
\begin{align}
\sup_{S_A\in\{[2^{S_p}]^{n}\setminus\{\emptyset\}\}} \left\{\lambda_{\max}^{\frac{1}{2}}\left(([A^r_i]_{S_A,:}[A^r_i]_{S_A,:}^T)^{-1}\right)\right\}\le c_M \label{eq.def.cM}
\end{align}
where $S_p=\{1,2,\ldots, n_{p}\}$, $2^{S_p}$ is the power set of $S_p$, and $[2^{S_p}]^n$ denotes the set of subsets of $2^{S_p}$ with cardinality not larger than $n$.

\begin{align}
\delta_i &\ge \frac{\delta-\bar{v}^cc_M|\breve{a}^{r}_i|}{1+c_M^2|\breve{a}^{r}_i|^2}\ge 
\frac{\delta-\bar{v}^cc_M \tau_a \delta^{-1} }{1+c_M^2\tau^2_a \delta^{-2}}
=\delta\frac{\delta^2-\bar{v}^cc_M \tau_a}{\delta^2+c_M^2\tau^2_a}
\label{eq.deltai.greaterthan1}
\end{align}
If
\begin{align}
\delta^2\ge 2\bar{v}^cc_M\tau_a+c_M^2\tau_a^2,
\end{align}
then \eqref{eq.deltai.greaterthan1} directly implies $\delta_i \ge \delta/2$.
This completes the proof of Property 3.
\end{IEEEproof}

Based on Proposition \ref{proposition.relaxation_parameter}, the following proposition shows a robust safety property of the controlled nominal systems in the presence of velocity-tracking errors.

\begin{proposition}\label{proposition.multi.robustISpSlyapunov}
Consider the mobile agent defined by \eqref{eq.multi.kinematicstransform}, and the controller defined by \eqref{eq.def.calP'_i}--\eqref{eq.def.v*_i.controller}. Given any class $\mathcal{K}_{\infty}$ function $\gamma^{\tilde{v}}_{V}$ and any positive constant $c_0$, choose
\begin{align}
\alpha_V(s) &= 4(1+c_0)\left(\gamma^{\tilde{v}}_{V}\right)^{-1}(s) \label{eq.proposition.alphaV.definition}
\end{align}
for $s\ge 0$. Then, $\alpha_V$ is of class $\mathcal{K}^e$, and the following properties hold:
\begin{enumerate}
\item For any $\tilde{p}_{ij} \neq 0$ with $i\in\mathcal{N}_a$ and $j\in\mathcal{N}_{ao}\setminus\{i\}$, the QP algorithm \eqref{eq.def.calP'_i}--\eqref{eq.chi.A_ia_i} is feasible and has a unique solution;
\item There exist $\beta_V \in \mathcal{KL}$, $\theta > 0$ and $d^{\tilde{v}}_{V}=(1+\theta)\mu_0$ such that for any $V_{ij}(0)\in\mathbb{R}_+$ and any piecewise continuous and bounded $\tilde{v}_i$, $\tilde{v}_j$ satisfying $\max_{k=i,j} \|\tilde{v}_k\|_t\le\left(\gamma^{\tilde{v}}_{V}\right)^{-1}(\mu(-D_s)-d^{\tilde{v}}_{V})$,
\begin{align}
V_{ij}(t)\le\beta_V(V_{ij}(0),t)+\gamma^{\tilde{v}}_{V}\Big(\max_{k=i,j} \|\tilde{v}_k\|_t\Big)+d^{\tilde{v}}_{V}\label{eq.proposition.multi.lyapunov.ISpS}
\end{align}
for all $0 \le t< \max\{T_i, T_j\}$.
\end{enumerate}
\end{proposition}

\begin{IEEEproof}
The properties are proved one-by-one.

{\bfseries Property 1:} For any $\tilde{p}_{ij} \neq 0$ with $i\in\mathcal{N}_a$ and $j\in\mathcal{N}_{ao}$, the feasible set $\mathcal{P}^{r}_i(p)$ defined in \eqref{eq.def.calP'_i.feasibleset} forms a convex, closed set \cite[Section 2.2.4]{Boyd-Vandenberghe-book-2004}. Since zero is always an element of $\mathcal{P}^{r}_i(p)$, the QP algorithm is feasible. The uniqueness of the solution is guaranteed by the projection theorem \cite[Proposition B.11]{Bertsekas-book-1997}.

{\bfseries Property 2:} Recall $\tilde{p}_{ij}=p_i-p_j$ in \eqref{eq.multi.kinematicstransform2} and $V_{ij}=V(\tilde{p}_{ij})$ in \eqref{eq.V_ij.simplify}. By taking the derivative of $V_{ij}$ along the trajectories of \eqref{eq.multi.kinematicstransform}, for any $[v^*_i;\delta_i]\in \mathcal{P}^{r}_{i}(p)$, if $V_{ij}\leq\mu(-D_s)$, we have
\begin{align}
&\nabla V_{ij}\dot{\tilde{p}}_{ij} =\nabla V_{ij}(v^*_i+\tilde{v}_i-v^*_j-\tilde{v}_j)\notag\\
%&~~~=-\alpha_\mu(V_{ij}) \hat{\tilde{p}}^{T}_{ij} (v^*_i-v^*_j+\tilde{v}_i-\tilde{v}_j)\notag\\
&~~~=\alpha_\mu(V_{ij})(-\hat{\tilde{p}}^{T}_{ij}(v^*_i-v^*_j)-\hat{\tilde{p}}^{T}_{ij}(\tilde{v}_i-\tilde{v}_j) )\notag\\
&~~~\le\alpha_{\mu}(V_{ij})\left(-\frac{\delta_{ij}}{\delta}\alpha_{V}(V_{ij}-\mu_0)-\hat{\tilde{p}}^{T}_{ij}(\tilde{v}_i-\tilde{v}_j)\right) \label{eq.Vij.derivative}
\end{align}
where $\hat{\tilde{p}}^{T}_{ij}=\tilde{p}^{T}_{ij}/|\tilde{p}_{ij}|$,
\begin{align}
\delta_{ij}=\begin{cases}
\delta_{i}+\delta_{j},&\text{for}~j\in\mathcal{N}_a\setminus \{i\},\\
\delta_{i},&\text{for}~j\in\mathcal{N}_o,
\end{cases}\label{eq.delta_ij.definition}
\end{align}
and 
\begin{align}
\alpha_{\mu}(s)&=
\begin{cases}
-\frac{\partial \mu(\mu^{-1}(s))}{\partial \mu^{-1}(s)},& \text{if}~s > 0,\\
0,& \text{if}~s=0.
\end{cases}\label{eq.def.alphamu}
\end{align}
For the inequality in \eqref{eq.Vij.derivative}, we used the constraint of the feasible set $\mathcal{P}^{s}_{i}(p)$ in \eqref{eq.def.QP.non_Lipschitz} and the property that $\mathcal{P}^{r}_i(p)\subseteq\mathcal{P}^s_i(p)$ given in Proposition \ref{proposition.calP_i.subset}.

With $\alpha_V$ defined by \eqref{eq.proposition.alphaV.definition}, for any $\tilde{p}_{ij}$ satisfying $V(\tilde{p}_{ij})\ge\mu_0$, it holds that
\begin{align}
-\alpha_V(V_{ij}-\mu_0) = -4(1+c_0)\left(\gamma^{\tilde{v}}_{V}\right)^{-1}(V_{ij}-\mu_0). \label{eq.alphaV.Vij.substitute}
\end{align}

From property 2 of Proposition \ref{proposition.relaxation_parameter}, for all $0 \le t< \max\{T_i, T_j\}$, it holds that
\begin{align}
\frac{\delta_{ij}}{\delta}\ge \frac{1}{2}.\label{eq.deltaij.ge.underlineDelta}
\end{align}

Then, \eqref{eq.Vij.derivative}, \eqref{eq.alphaV.Vij.substitute} and \eqref{eq.deltaij.ge.underlineDelta} together imply that
\begin{align}
\nabla V_{ij}\dot{\tilde{p}}_{ij}\le& 2\alpha_\mu(V_{ij})\big(-(1+c_0)(\gamma^{\tilde{v}}_{V})^{-1}(V_{ij}-\mu_0)\notag \\
&+\max_{k=i,j} \|\tilde{v}_k\|_t\big)\label{eq.lamma4.dotV.res0}
\end{align}
as long as $\mu_0\le V_{ij}\le\mu(-D_s)$.

Denote $d_V^{\tilde{v}}=(1+\theta)\mu_0$. In the case of
\begin{align}
\gamma^{\tilde{v}}_{V}\Big(\max_{k=i,j} \|\tilde{v}_k\|_t\Big)+d^{\tilde{v}}_{V}\le V_{ij}\le\mu(-D_s), \label{eq.proposition4.V.cond}
\end{align}
it can be easily verified that
\begin{align}
V_{ij}-\mu_0\ge\gamma^{\tilde{v}}_{V}\Big(\max_{k=i,j} \|\tilde{v}_k\|_t\Big),
V_{ij}-\mu_0\ge\frac{\theta}{1+\theta}V_{ij}.\label{eq.Vm.tildevm_Vm.relationship}
\end{align}
By substituting \eqref{eq.Vm.tildevm_Vm.relationship} into \eqref{eq.lamma4.dotV.res0}, we have that
\begin{align}
\nabla V_{ij}\dot{\tilde{p}}_{ij}
\le-2c_0 \alpha_{\mu}(V_{ij}) (\gamma^{\tilde{v}}_{V})^{-1}\left(\frac{\theta}{1+\theta}V_{ij}\right)\label{eq.lamma4.dotV.res}
\end{align}
holds in the case of \eqref{eq.proposition4.V.cond}.

Based on the discussion above, if $\alpha_{\mu}\in\mathcal{K}$, then there exists a $\beta_V\in \mathcal{KL}$ for property \eqref{eq.proposition.multi.lyapunov.ISpS} \cite{Sontag-2008}.

Now, we show $\alpha_{\mu}\in\mathcal{K}$. By using the properties of $\mathcal{M^-C}$ functions and properties of limits of composition functions \cite[Section 8.16]{Binmore-book-1982}, we have $\lim_{s\rightarrow \infty}\mu(s) = 0$, $\lim_{s\rightarrow 0}\mu^{-1}(s) = \infty$, $\lim_{s\rightarrow \infty} \partial \mu(s)/ \partial s = 0$, and $\lim_{s\rightarrow 0} - \partial \mu(\mu^{-1}(s))/ \partial \mu^{-1}(s) = 0$. This validates the right continuity of $\alpha_{\mu}$ at the origin. Since $\mu$ is strictly decreasing, for any $s_2>s_1>0$, it holds that $\mu^{-1}(s_1)>\mu^{-1}(s_2)>0$. Also recall that $\mu$ is strictly convex. By using \cite[Equation 3.3]{Boyd-Vandenberghe-book-2004}, we have
%$\mu(\mu^{-1}(s_1))>\mu(\mu^{-1}(s_2))-\alpha_\mu(s_2)(\mu^{-1}(s_1)-\mu^{-1}(s_2))$ and $\mu(\mu^{-1}(s_2))>\mu(\mu^{-1}(s_1))-\alpha_\mu(s_1)(\mu^{-1}(s_2)-\mu^{-1}(s_1))$.
\begin{align}
\mu(\mu^{-1}(s_1))&>\mu(\mu^{-1}(s_2))-\alpha_\mu(s_2)(\mu^{-1}(s_1)-\mu^{-1}(s_2)), \\
\mu(\mu^{-1}(s_2))&>\mu(\mu^{-1}(s_1))-\alpha_\mu(s_1)(\mu^{-1}(s_2)-\mu^{-1}(s_1)). 
\end{align}
Then, it can be verified that
\begin{align}
\alpha_\mu(s_2)>\frac{s_2-s_1}{\mu^{-1}(s_1)-\mu^{-1}(s_2)}>\alpha_\mu(s_1),
\end{align}
and thus $\alpha_{\mu}$ is strictly increasing. This proves $\alpha_{\mu} \in \mathcal{K}$. 
%
%Then, it can be proved that $\alpha_{\mu\theta}$ defined in \eqref{eq.def.alphamuepsilon} is of class $\mathcal{K}_{\infty}$.
\end{IEEEproof}

\begin{remark}
Proposition \ref{proposition.multi.robustISpSlyapunov} employs the gain $\gamma_{V}^{\tilde{v}}$ to represent the influence of $\max\{|\tilde{v}_i|, |\tilde{v}_j|\}$ on $V(\tilde{p}_{ij})$, and accordingly, the robustness of $V(\tilde{p}_{ij})$ with respect to $\max\{|\tilde{v}_i|, |\tilde{v}_j|\}$.
With $v^*_i$ considered as the control input, Proposition \ref{proposition.multi.robustISpSlyapunov} also shows how $\alpha_V$ can be chosen for a desired gain $\gamma_{V}^{\tilde{v}}$.
\end{remark}

\subsection{Response of the Uncertain Actuation System}\label{subsection.actuationresponse}

In this subsection, we study the dynamic response of the actuation system with the velocity reference signal generated by the refined QP-based controller.
%Specifically, we give the following proposition to show how $V_m$ influences $\tilde{v}_i$.

\begin{proposition}\label{proposition.QPsolution.multi}
Under Assumptions \ref{assumption.multi.referencetracking} and \ref{assumption.multi.boundedprimarycontrol}, consider the multi-agent system modeled by \eqref{eq.def.p_i.kinematics}--\eqref{eq.def.v_i.actuation} and the controller defined by \eqref{eq.def.calP'_i}--\eqref{eq.def.v*_i.controller}. Then, the following properties hold:
\begin{enumerate}
\item For $t\in[0, T_i)$, the solution to the QP algorithm \eqref{eq.def.calP'_i}--\eqref{eq.chi.A_ia_i} is Lipschitz with respect to $a^r_i$ and $v^c_i$;
\item There exist a class $\mathcal{K}$ function $\alpha^{V}_{v^*}$ such that
\begin{align}
|v^*_i(t)|&\le\alpha^{V}_{v^*}(V_R(t))+\bar{v}^c,\label{eq.vset.multi.upper}
\end{align}
for all $t\ge 0$, where $V_R$ is defined in \eqref{eq.V.running.definition};
\item There exist $\beta^V_{\tilde{v}}\in\mathcal{KL}$, $\gamma^{V}_{\tilde{v}}$, $\gamma^{\tilde{v}}_{\tilde{v}}\in\mathcal{K}$, and a constant $d^V_{\tilde{v}}\in\mathbb{R}_+$ such that for all $z_{i}(0)\in\mathbb{R}^m$,
\begin{align}
&|\tilde{v}_i(t)|\le\notag\\
&\begin{cases}
\max\left\{\hspace{-3pt} \begin{array}{l}
\beta_{\tilde{v}}^V(|z_{i}(0)|, t) + \gamma^{V}_{\tilde{v}}(\|V_R\|_t)+d^{V}_{\tilde{v}}, \hspace{-3pt}\\
\gamma^{\tilde{v}}_{\tilde{v}}(\|\tilde{v}_m\|_t)
\end{array}\right\},\\                       \text{\hspace{103pt}for }t\in[0,T_i),\\
c_{\tilde{v}}e^{-\lambda (t-T_i)}|z_i(T_i)|, \text{\hspace{20pt}for }t\in[T_i,\infty).
\end{cases}\hspace{-18pt}\label{eq.solution.derivative.boundedness}
\end{align}
\end{enumerate}
\end{proposition}

\begin{IEEEproof}
Before the proofs, we rewrite the QP algorithm defined by \eqref{eq.def.calP'_i}--\eqref{eq.chi.A_ia_i} as
\begin{align}
u=\argmin_{u\in\{x\in\mathbb{R}^{n+1}:M_Fx\le 0,[x]_{n+1,:}\ge 0\}} 0.5u^Tu+ru
\end{align}
where
\begin{align}
u=[v^*_i; \delta_i],~~M_F= [A^r_i, a^r_i],~~r=-[v^c_i; \delta].\label{eq.multi.conditions}
\end{align}
As for the proof of property 3 of Proposition \ref{proposition.relaxation_parameter}, we define $M_i=[\breve{A}^r_i,\breve{a}^r_i]$ to represent the non-redundant active constraints of the QP algorithm, where $\breve{A}^r_i$ and $\breve{a}^r_i$ are submatrices of $A^r_i$ and $a^r_i$, respectively.

Now we prove properties of Proposition \ref{proposition.QPsolution.multi} one-by-one.

{\bfseries Property 1:} For $t\in [0, T_i)$, we prove the Lipschitz continuity of the solution to the QP algorithm by using \cite[Theorem 3.1]{Hager-SIAMControl-1979} as follows.
\begin{itemize}
\item As already proved for property 1 of Proposition \ref{proposition.multi.robustISpSlyapunov}, the QP algorithm has a unique solution.
\item Since $\breve{a}^r_i$ is a submatrix of $a^r_i$, according to property 2 of Proposition \ref{proposition.relaxation_parameter}, $\breve{a}^r_i$ is bounded; since $\breve{A}^r_i$ is composed of unit vectors, $M_i$ is bounded.
\item We use $\breve{n}$ to represent the number of non-redundant active constraints of the QP algorithm. By using the definition of $T_i$, we have $\breve{n} \le n$ for all $t\in [0,T_i)$. Since any $n$ rows of $A^r_i$ are linearly independent and $\breve{n}\leq n$, $\breve{A}^r_i$ and $M_i$ are full row rank, which means that there exists a $c_{\lambda}>0$ such that $\lambda_{\min}^{1/2}(\breve{A}^r_i\breve{A}^{rT}_i)\ge c_{\lambda}$. It can be directly verified that $|M_i^T\lambda|\ge\lambda_{\min}^{1/2}(M_iM_i^T)|\lambda|\ge \lambda_{\min}^{1/2}(\breve{A}^r_i\breve{A}^{rT}_i)|\lambda|\ge c_{\lambda}|\lambda|$ for all $\lambda$.
\end{itemize}

Then, with all the conditions required by \cite[Theorem 3.1]{Hager-SIAMControl-1979} satisfied, we can prove the Lipschitz continuity of the solution with respect to $M_F$ and $r$. Since $\delta$ and $A^r_i$ are constant, property 1 can be proved by recalling the definitions of $M_F$ and $r$ in \eqref{eq.multi.conditions}.

{\bfseries Property 2:} When $t\ge T_i$, the controller \eqref{eq.def.v*_i.controller} guarantees that $v^*_i \equiv 0$. 
When $t\in[0, T_i)$, we consider the cases of $[v^c_i; \delta]\notin\mathcal{P}^{r}_i$ and $[v^c_i;\delta]\in\mathcal{P}^{r}_i$ separately. 
In the case of  $[v^c_i; \delta]\notin\mathcal{P}^{r}_i$, we have $v^*_i = v^c_i$. 

Now we consider the case of $[v^c_i; \delta] \notin\mathcal{P}^{r}_i$. Define
\begin{align}
q_i(v^c_i) = \argmin\limits_{u_i\in \{x\in\mathbb{R}^n: A^r_ix+a^r_i\delta_i \le 0\} } 0.5 u^T_iu_i - v^{cT}_i u_i.\label{eq.QP.temp}
\end{align}
We use $\check{A}^r_i$ and $\check{a}^r_i\delta_i$ to represent the non-redundant active constraints of \eqref{eq.QP.temp}. Clearly, $\check{A}^r_i$ and $\check{a}^r_i\delta_i$ are submatrices of $A^r_i$ and $a^r_i\delta_i$, respectively.
By using the projection theorem \cite[Proposition B.11]{Bertsekas-book-1997}, $q_i(v^c_i)$ is continuous and nonexpansive, which means that
\begin{align}
|q_i(v^c_i)-q_i(0)| \le |v^c_i|.
\end{align}
From \eqref{eq.QP.temp}, we also have $v^*_i = q_i(v^c_i)$. Then following \cite[Example 2.1.5]{Bertsekas-book-1997}, we have
\begin{align}
|v^*_i| &\le |v^c_i| + |q_i(0)| 
= \bar{v}^c + \left|\check{A}^{rT}_i(\check{A}^r_i\check{A}^{rT}_i)^{-1}\check{a}^{r}_i\delta_i\right| \notag \\
&\le \bar{v}^c + \delta_i \sqrt{\check{a}^{rT}_i(\check{A}^r_i\check{A}^{rT}_i)^{-1}\check{a}^{r}_i}. \label{eq.continuous_nonexpansive}
\end{align}
By applying the Lagrange multiplier algorithms and Karush-Kuhn-Tucker optimality conditions \cite{Bertsekas-book-1997}, we have
\begin{align}
\check{a}^{rT}_i(\check{A}^r_i\check{A}^{rT}_i)^{-1} \ge 0. \label{eq.lagrange.multiplier}
\end{align}
Denote $\bar{a}_i = \max\{\check{a}^r_i\} 1_{n\times 1}$. It can be easily verified that
\begin{align}
(\bar{a}_i-\check{a}^{r}_i)^T(\check{A}^r_i\check{A}^{rT}_i)^{-1}(\bar{a}_i-\check{a}^{r}_i) \ge 0,  \\
\check{a}^{rT}_i(\check{A}^r_i\check{A}^{rT}_i)^{-1}(\bar{a}_i-\check{a}^{r}_i) \ge 0,
\end{align}
and thus, 
\begin{align}
&\bar{a}_i^T(\check{A}^r_i\check{A}^{rT}_i)^{-1}\bar{a}_i = (\bar{a}_i-\check{a}^{r}_i)^T(\check{A}^r_i\check{A}^{rT}_i)^{-1}(\bar{a}_i-\check{a}^{r}_i)  \notag \\
&~~~~~~~~ + 2\check{a}^{rT}_i(\check{A}^r_i\check{A}^{rT}_i)^{-1}(\bar{a}_i-\check{a}^{r}_i) + \check{a}^{rT}_i(\check{A}^r_i\check{A}^{rT}_i)^{-1}\check{a}^r_i \notag \\
&~~~~~~~~ \ge \check{a}^{rT}_i(\check{A}^r_i\check{A}^{rT}_i)^{-1}\check{a}^r_i, \label{eq.activeconstraint.upperbound}
\end{align}
where $\bar{a}_i = \max\{\check{a}^r_i\} 1_{n\times 1}$. This, together with \eqref{eq.continuous_nonexpansive} implies
\begin{align}
&| v^*_i|  \le |v^c_i| + \delta_i\sqrt{\bar{a}_i^T(\check{A}^r_i\check{A}^{rT}_i)^{-1}\bar{a}_i} \le \bar{v}^c + c_M\sqrt{n} \max\{\check{a}^r_i\delta\} \notag \\
%&\le \bar{v}^c + \lambda_{\max}^{\frac{1}{2}}((\check{A}^r_i\check{A}^{rT}_i)^{-1}) \delta^{-1} \delta_i \sqrt{n} \max\{\check{a}^r_i\delta\} \notag \\
%&\le \bar{v}^c + c_M\sqrt{n} \max\{\check{a}^r_i\delta\} \notag \\
&\le \bar{v}^c + c_M\sqrt{n} \max_{\begin{subarray}{l} j\in\{1,\ldots,n_p:[A^r_i]_{j,:} \text{~is a row of~} \check{A}^r_i\}\\
k\in\{1,\ldots,n_{ao}-1 \}
\end{subarray}} [\varphi(A_i^s,a^s_i\delta,c_P)]_{j,k} \notag \\
&\le \bar{v}^c + c_M\sqrt{n} \alpha_V(V_R - \mu_0) \le c_M\sqrt{n} \alpha_V(V_R) + \bar{v}^c, \label{eq.v^*.BIBO}
\end{align}
where we use \eqref{eq.activeconstraint.upperbound} for the first inequality, use the definition of $c_M$ in \eqref{eq.def.cM} and property 3 of Proposition \ref{proposition.relaxation_parameter} for the second inequality and use the equivalent representation of $\varphi$ in \eqref{eq.barkbara.chijk} for the fourth inequality. 

Thus, property 2 is proved by defining $\alpha^V_{v^*}(s)=c_M\sqrt{n} \alpha_V(s)$.

{\bfseries Property 3:}
When $t\ge T_i$, controller \eqref{eq.def.v*_i.controller} gives $v^*_i \equiv 0$. Then, by using the definition of $\tilde{v}_i$ in \eqref{eq.multi.referencetrackingerror} and property \eqref{eq.asmp.vtilde_i.IOpS}, we have
\begin{align}
|\tilde{v}_i(t)| \le c_{\tilde{v}}e^{-\lambda(t-T_i)}|z_i(T_i)| \label{eq.def.tildev_case2}
\end{align}
for all $t\ge T_i$.

Now, we consider the case of $t\in [0, T_i)$. Property 1 already shows the Lipschitz continuity of \eqref{eq.def.v*_i.controller} with respect to $a^r_i$ and $v^c_i$. 
To prove property 3, we first show the existence of two class $\mathcal{K}$ functions $\alpha^{V}_{v^{*d}}$ and $\alpha^{\tilde{v}}_{v^{*d}}$, and a positive constant $c^V_{v^{*d}}$ such that
\begin{align}
|v^{*d}_i(t)| &\le \alpha^{\tilde{v}}_{v^{*d}}(|\tilde{v}_m(t)|) + \alpha^{V}_{v^{*d}}(V_R(t)) + c^V_{v^{*d}}\label{eq.multi.v*d.traj.limit}
\end{align}
for all $t\in[0, T_i)$.
By using weak triangular inequality in \cite{Jiang-Teel-Praly-MCSS-1994}, properties \eqref{eq.asmp.vtilde_i.IOpS} and \eqref{eq.multi.v*d.traj.limit} together guarantee the first case in \eqref{eq.solution.derivative.boundedness} with 
\begin{align}
&d^{V}_{\tilde{v}}= (1+k_{\tilde{v}})\gamma^{v^{*d}}_{\tilde{v}} \circ \rho^{v^{*d}}_{\tilde{v}} (c^V_{v^{*d}}), \label{eq.def.dV2vtilde} \\
&\gamma^V_{\tilde{v}}(s) = (1+k_{\tilde{v}})\gamma^{v^{*d}}_{\tilde{v}} \circ \rho^{v^{*d}}_{\tilde{v}}(\alpha^V_{v^{*d}}(s) + c^V_{v^{*d}}) - d^{V}_{\tilde{v}}, \label{eq.def.gammaV2vtilde} \\
&\beta_{\tilde{v}}^V(s, t)= (1+k_{\tilde{v}})\beta_{\tilde{v}}(s, t), \label{eq.def.betaV2vtilde} \\
&\gamma^{\tilde{v}}_{\tilde{v}}(s) = (1+k_{\tilde{v}}^{-1})\gamma^{v^{*d}}_{\tilde{v}}\!\circ\!\rho^{v^{*d}}_{\tilde{v}}\!\circ\!(\rho^{v^{*d}}_{\tilde{v}}\!\!\!\!-\id)^{-1}\!\circ\!\alpha^{\tilde{v}}_{v^{*d}}(s),\hspace{-5pt}
\label{eq.def.gammavtilde2vtilde}
\end{align}
where $\rho^{v^{*d}}_{\tilde{v}}$ is a function of class $\mathcal{K}_{\infty}$ such that $\rho^{v^{*d}}_{\tilde{v}}-\id$ is of class $\mathcal{K}_{\infty}$.

In the case of $[v^c_i; \delta] \in \mathcal{P}^{r}_i$, $v^c_i$ is the solution to the QP algorithm, and thus $|v^{*d}_i| \le \bar{v}^{cd}$. In this case, property \eqref{eq.multi.v*d.traj.limit} is obvious.

Now we consider the case of $[v^c_i; \delta] \notin\mathcal{P}^{r}_i$.
By using \cite[Theorem 3.1]{Hager-SIAMControl-1979}, there exists a positive constant $c_{\mathcal{P}} \ge 1$ such that
%\begin{align}
%\left|v^{*d}_i\right| &\le \left|[v^{*d}_i; \delta^d_i]\right| \notag \\
%&\le c_{\mathcal{P}} \bar{v}^{cd} + 2c_{\mathcal{P}}^2\sqrt{\bar{v}^{c^2}+\delta^2}\left| D^+a^r_i(p(t)) \right|. \label{eq.def.v*d_i.hager}
%\end{align}
\begin{align}
\left|v^{*d}_i\right| \le c_{\mathcal{P}} \bar{v}^{cd} + 2c_{\mathcal{P}}^2\sqrt{\bar{v}^{c^2}+\delta^2}\left| D^+a^r_i(p(t)) \right|. \label{eq.def.v*d_i.hager}
\end{align}
If there exist two class $\mathcal{K}$ functions $\alpha^{V}_{\varphi}$ and $\alpha^{\tilde{v}}_{\varphi}$ such that
\begin{align}
\left| D^+a^r_i(p(t)) \right| \le \alpha^{V}_{\varphi}(V_R) + \alpha^{\tilde{v}}_{\varphi}(|\tilde{v}_m|), \label{eq.chi.bounded}
\end{align}
then property \eqref{eq.multi.v*d.traj.limit} is established.

Now we prove inequality \eqref{eq.chi.bounded} holds for all $[v^c_i; \delta] \notin\mathcal{P}^{r}_i$.
By taking derivatives of $A^s_i$ and $a^s_i$ with respect to $t$, we obtain
\begin{align}
\frac{\partial A^s_i}{\partial t} &= 
\left[A^d_{i1}, \ldots, A^d_{i(i-1)}, A^d_{i(i+1)},\ldots, A^d_{in_{ao}} \right]^T \label{eq.partialAi_t}\\
\frac{\partial a^s_i}{\partial t} &= 
\left[a^d_{i1}, \ldots, a^d_{i(i-1)}, a^d_{i(i+1)},\ldots, a^d_{in_{ao}} \right]^T \label{eq.partialai_t}
\end{align}
where $A^d_{ij}=-(I-\hat{\tilde{p}}_{ij}\hat{\tilde{p}}^{T}_{ij} )(v_i-v_j)|\tilde{p}_{ij}|^{-1}$ and 
$a^d_{ij}=\delta^{-1} \alpha_{\mu}(V_{ij}) \hat{\tilde{p}}^{T}_{ij} (v_i-v_j)(\partial \alpha_V(V_{ij}-\mu_0))/(\partial(V_{ij}-\mu_0))$ with $j \in \mathcal{N}_{ao} \setminus\{i\}$.
By using the definition of velocity-tracking error in \eqref{eq.multi.referencetrackingerror}, the maximal velocity of agents satisfies
\begin{align}
\max_{i\in\mathcal{N}_a}|v_i| 
&= |\tilde{v}_m| + \max_{i\in\mathcal{N}_{a}} |v^*_i| 
= |\tilde{v}_m| + \max_{i\in\mathcal{N}_{a}\setminus \mathcal{N}_s} |v^*_i| 
\notag \\
&\le |\tilde{v}_m| + \alpha^V_{v^*}(V_R) + \bar{v}^c.\label{eq.vm.limit}
\end{align}
The definition of $\alpha_V$ shows the lower bound of $\alpha_V(V_{ij}-\mu_0)$ and implies that there exist a class $\mathcal{K}$ function $\alpha^d_{V}$ and a constant $c^{d}_V \in \mathbb{R}_+$ such that
\begin{align}
\left|\frac{\partial \alpha_V(V_{ij}-\mu_0)}{\partial (V_{ij}-\mu_0)}\right| \le \alpha^d_{V}(V_R)+c^{d}_V. \label{eq.alphaV.partial}
\end{align}
Then, \eqref{eq.partialAi_t}, \eqref{eq.partialai_t} and \eqref{eq.alphaV.partial} imply
\begin{align}
\left|\frac{\partial A^s_i}{\partial t} \right| &\le 2\sqrt{n_{ao}-1}\alpha_p(V_R)\max_{i\in\mathcal{N}_a}|v_i|, \label{eq.proof.ieq.dA_i}\\
\left|\frac{\partial a^s_i}{\partial t} \right| &\le \frac{2\sqrt{n_{ao}-1}}{\delta} \alpha_{\mu}(V_R)(\alpha^d_{V}(V_R)+c^{d}_V)\max_{i\in\mathcal{N}_a}|v_i|, \!\!\label{eq.proof.ieq.da_i}
\end{align}
where $\alpha_{\mu}$ is a class $\mathcal{K}$ function defined in \eqref{eq.def.alphamu}, and 
\begin{align}
\alpha_p(s) =\begin{cases}
\frac{1}{\mu^{-1}(s) +D_s}, & \text{for}~s > 0, \\
0, & \text{for}~s = 0.
\end{cases}\label{eq.alpha_p}
\end{align}
Clearly, $\alpha_p(s)$ is a class $\mathcal{K}$ function. 
%We have $\alpha_p(V(\tilde{p})) =|\tilde{p}|^{-1}$ by recalling the definitions of $\mu$ and $V(\tilde{p})$.
Each row of $A^s_i$ is a unit vector, and $a^s_i$ is bounded when $[v^{c}_i; \delta] \notin \mathcal{P}^{r}_i$ and $t\in [0, T_i)$ (see \eqref{eq.a^s_i.bounded}).
By using the definition of $\varphi$ in \eqref{eq.chi.A_ia_i}, we have that $\varphi$ is locally Lipschitz with respect to $A^s_i$ and $a^s_i$, and thus, there exists a constant $c_{\varphi} > 0$ satisfying
\begin{align}
\left| D^+a^r_i(p(t)) \right| 
&\le c_{\varphi}\left(\left|\frac{\partial A^s_i}{\partial t}\right| + \left|\frac{\partial a^s_i}{\partial t}\right| \right)
\end{align}
for all $t\in [0, T_i)$. By substituting \eqref{eq.vm.limit}, \eqref{eq.proof.ieq.dA_i} and \eqref{eq.proof.ieq.da_i} into the inequality above, we can prove that \eqref{eq.chi.bounded} holds with
\begin{align}
\alpha^{V}_{\varphi}(s) &= \left(\alpha^V_{v^*}(s) + \bar{v}^c + \frac{\alpha^d_{Aa}(s)}{4k_{\varphi}} \right) \alpha^d_{Aa}(s), \\
\alpha^{\tilde{v}}_{\varphi}(s) &= k_{\varphi}s^2,\label{eq.def.alpha_V_chi}
\end{align}
where $k_{\varphi}$ is a positive constant and $\alpha^d_{Aa}(s) = 2c_{\varphi}(n_{ao}-1)^{1/2}\left(\alpha_p(s) +\delta^{-1} \alpha_{\mu}(s)(\alpha^d_{V}(s)+c^{d}_V)\right)$. From \eqref{eq.def.v*d_i.hager}, \eqref{eq.chi.bounded} and \eqref{eq.def.alpha_V_chi}, we have that \eqref{eq.multi.v*d.traj.limit} holds for all $t\in[0, T_i)$.
This together with \eqref{eq.def.tildev_case2} proves property 3.

This ends the proof of Proposition \ref{proposition.QPsolution.multi}.
\end{IEEEproof}

\subsection{Small-Gain Analysis for Safety of the Multi-Agent System}
\label{subsection.smallgain}

\begin{figure}[h!]
\centering
\begin{tikzpicture}[scale=1]
\node at (2,1.5) {};
\node[fill = white, draw = black, circle, minimum size=1.35cm] (ua) at (22:4.0204) {$\tilde{v}_m$};
\node[fill = white, draw = black, circle, minimum size=1.35cm] (vt) at (80:1.5221) {$V_m$};
\node at (2,0.2) {$\gamma^{\tilde{v}}_{V}$};
\draw[-latex] (3.3121,0.9906) arc (-49:-131:2);
\node at (2,2.8) {$\gamma^{V}_{\tilde{v}}$};
\draw[-latex] (0.7144,2.0321) arc (130.0003:49:2);
\node at (5.75,1.5) {$\gamma^{\tilde{v}}_{\tilde{v}}$};
\draw[-latex] (3.9636,0.8636) arc (-135:136:0.9);
\end{tikzpicture}
\caption{Gain interconnection between $V_m$ and $\tilde{v}_m$.}
\label{figure.Small gain diagram 2}
\end{figure}
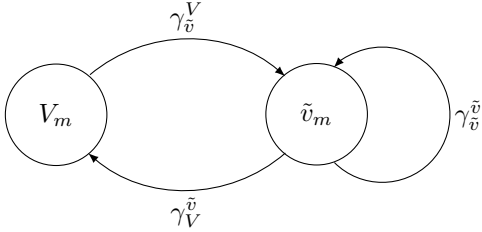

Propositions \ref{proposition.multi.robustISpSlyapunov} and \ref{proposition.QPsolution.multi} naturally result in two interconnected subsystems, each of which admits a gain property (see Figure \ref{figure.Small gain diagram 2}). We employ a small-gain analysis to guarantee the safety of the closed-loop system.

We suppose that the parameters are chosen to satisfy the following small-gain-like condition:
\begin{align}
&(\id + \epsilon_1)\circ\gamma^{\tilde{v}}_{V} \circ(\id + \epsilon_2) \circ \gamma^V_{\tilde{v}}(s) \le s, \notag \\ 
& \hspace{75pt} \forall s \in [\mu(0), \mu(D-D_s)] \label{eq.smallgain.multi}\\
&\gamma^{\tilde{v}}_{\tilde{v}}(s) < s, \hspace{30pt}\forall s\in[d^{V}_{\tilde{v}},\alpha_V(\mu(D-D_s)-\mu(0))] \label{eq.vtilde.convergence}
\end{align}
with $\epsilon_1,\epsilon_2\in\mathcal{K}_{\infty}$.

For any agent $i\in\mathcal{N}_a$ and any agent or obstacle $j\in\mathcal{N}_{ao}\setminus \{i\}$, we consider the following two cases.

{\bfseries Case (A): $t < \min_{k\in \mathcal{N}_a} T_k$}.

Recall the definitions of $V_m$, $\tilde{v}_m$ and $z_m$ given before Theorem \ref{theorem.smallgaincondition2}. In this case, properties \eqref{eq.proposition.multi.lyapunov.ISpS} and \eqref{eq.solution.derivative.boundedness} imply
\begin{align}
|\tilde{v}_m(t)|&\le\max\left\{\begin{array}{l}
\beta_{\tilde{v}}^V(|z_m(0)|, t) + \gamma^{V}_{\tilde{v}}(\|V_m\|_t)+d^{V}_{\tilde{v}},
\\
\gamma^{\tilde{v}}_{\tilde{v}}(\|\tilde{v}_m\|_t)
\end{array}
\right\},\\
V_m(t)&\le\beta_V(V_m(0), t) + \gamma^{\tilde{v}}_{V} (\|\tilde{v}_m\|_t) + d^{\tilde{v}}_{V}.
\end{align}

By taking the superemum of the left-hand sides of the inequalities above and using the definition of class $\mathcal{KL}$ functions, we have
\begin{align}
\|\tilde{v}_m\|_t&\le\max\left\{\begin{array}{l}
\beta_{\tilde{v}}^V(|z_m(0)|, 0) + \gamma^{V}_{\tilde{v}}(\|V_m\|_t)+d^{V}_{\tilde{v}},
\\
\gamma^{\tilde{v}}_{\tilde{v}}(\|\tilde{v}_m\|_t)
\end{array}
\right\},\label{proof.theorem.main.1}\\
\|V_m\|_t&\le\beta_V(V_m(0), 0) + \gamma^{\tilde{v}}_{V} (\|\tilde{v}_m\|_t) + d^{\tilde{v}}_{V}.\label{proof.theorem.main.2}
\end{align}

With $\gamma_{\tilde{v}}^{\tilde{v}}(s)<s$ for $d^{V}_{\tilde{v}}\le s\le \alpha_V(\mu(D-D_s)-\mu(0))$ given by \eqref{eq.vtilde.convergence}, property \eqref{proof.theorem.main.1} implies
\begin{align}
\|\tilde{v}_m\|_t&\le\beta_{\tilde{v}}^V(|z_m(0)|, 0) + \gamma^{V}_{\tilde{v}}(\|V_m\|_t)+d^{V}_{\tilde{v}}\label{proof.theorem.main.3}
\end{align}
as long as $\|\tilde{v}_m\|_t\leq\alpha_V(\mu(D-D_s)-\mu(0))$.

By substituting \eqref{proof.theorem.main.2} into \eqref{proof.theorem.main.3}, we have
\begin{align}
\|V_m\|_t \le \gamma^{\tilde{v}}_{V} \circ (\id + \epsilon_2) \circ \gamma^{V}_{\tilde{v}}(\|V_m\|_t) + \Delta_V\label{proof.theorem.main.4}
\end{align}
where
\begin{align}
\Delta_V&= \gamma^{\tilde{v}}_{V} \circ (\id +\epsilon_2^{-1})\circ (\beta^{V}_{\tilde{v}}(|z_m(0)|, 0)+ d^{V}_{\tilde{v}}) \notag \\
& ~~~+\beta_V(V_m(0), 0) + d^{\tilde{v}}_{V}
\end{align}
with $\epsilon_2$ being a class $\mathcal{K}_{\infty}$ function.

Condition \eqref{eq.smallgain.multi} implies
\begin{align}
\gamma^{\tilde{v}}_{V} \circ(\id + \epsilon_2) \circ \gamma^V_{\tilde{v}}(s) \le (\id + \epsilon_1)^{-1}(s)
\end{align}
for all $s \in [\mu(0), \mu(D-D_s)]$, and thus
\begin{align}
\gamma^{\tilde{v}}_{V} \circ(\id + \epsilon_2) \circ \gamma^V_{\tilde{v}}(s) \le (\id + \epsilon_1)^{-1}(s)+d_1\label{proof.theorem.main.5}
\end{align}
for all $s \in [0, \mu(D-D_s)]$, with $d_1=\gamma^{\tilde{v}}_{V} \circ(\id + \epsilon_2) \circ \gamma^V_{\tilde{v}}(\mu(0))$.

By substituting \eqref{proof.theorem.main.5} into \eqref{proof.theorem.main.4}, we have
\begin{align}
\|V_m\|_t \le (\id + \epsilon_1)^{-1}(\|V_m\|_t)+d_1 + \Delta_V, \label{proof.theorem.main.6}
\end{align}
and thus
\begin{align}
\|V_m\|_t &\le (\id -(\id +\epsilon_1)^{-1})^{-1}(d_1 + \Delta_V)\nonumber\\
&=(\id +\epsilon_1^{-1})(d_1 + \Delta_V),
\end{align}
provided that $\|V_m\|_t\leq\mu(D-D_s)$.
With \eqref{proof.theorem.main.3}, this means that
\begin{align}
\|\tilde{v}_m\|_t&\le\beta_{\tilde{v}}^V(|z_m(0)|, 0)\nonumber\\
&~~~+ \gamma^{V}_{\tilde{v}}\circ(\id +\epsilon_1^{-1})(d_1 + \Delta_V)+d^{V}_{\tilde{v}}. \label{eq.running.trackingerror}
\end{align}

It can be verified that, $\|V_m\|_t$ and $\|\tilde{v}_m\|_t$ are monotone with respect to $\mu(0)$, $\bar{v}^{cd}$ and the upper bounds of the initial states and the external inputs. Thus, all the conditions in the proof above can be satisfied by appropriately choosing these values, and the boundedness of $V_m$ and $\tilde{v}_m$ is proved. 
Then, by directly applying properties \eqref{eq.vset.multi.upper} and \eqref{eq.asmp.z_i.ISS}, the maximal state $z_m$ of the actuation systems is bounded.

{\bfseries Case (B): $t \ge \min_{k\in \mathcal{N}_a} T_k$.}

In this case, $\mathcal{N}_S$ is not empty and the cardinality of $\mathcal{N}_S$ is nondecreasing with respect to $t$.

In accordance with the definitions of $V_R$ and $V_S$ in \eqref{eq.V.running.definition} and \eqref{eq.V.breaking.definition}, we define 
\begin{align}
\tilde{v}_R=\argmax_{x\in\{\tilde{v}_i:i\in\mathcal{N}_a\setminus\mathcal{N}_S\}} |x|,~~~
\tilde{v}_S=\argmax_{x\in\{\tilde{v}_i:i\in\mathcal{N}_S\}} |x|,
\end{align}
to represent the maximal velocity-tracking error of the agents belonging to $\mathcal{N}_a\setminus\mathcal{N}_S$ and $\mathcal{N}_S$, respectively. By using the definitions of $V_m$, $V_R$, $V_S$, $\tilde{v}_m$, $\tilde{v}_R$ and $\tilde{v}_S$, we have
\begin{align}
V_m = \max\{V_R, V_S\}, ~~~~ |\tilde{v}_m| = \max\{|\tilde{v}_R|, |\tilde{v}_S|\}.
\end{align}
Thus, the boundedness of $V_m$ and $\tilde{v}_m$ is guaranteed by proving the boundedness of $V_R$, $V_S$, $\tilde{v}_R$ and $\tilde{v}_S$.

Denote
\begin{align}
T_S=\max_{i\in\mathcal{N}_S} T_i.
\end{align}

{\bfseries (B1) Boundedness of $\tilde{v}_S$ and $V_S$.} For any $i\in\mathcal{N}_S$ and $t\ge T_S$,  we have $v^{*d}_i(t)\equiv 0$, and the following property can be verified by directly applying property \eqref{eq.asmp.vtilde_i.IOpS}:
\begin{align}
|\tilde{v}_i(t)| \le c_{\tilde{v}}|z_i(T_S)|e^{-\lambda(t-T_S)},
\end{align}
which implies
\begin{align}
|\tilde{v}_S(t)| \le c_{\tilde{v}}|z_m(T_S)|e^{-\lambda(t-T_S)} \le c_{\tilde{v}}|z_m(T_S)|.\label{eq.the.vtildes.bound1}
\end{align}
This means the boundedness of $\tilde{v}_S$.

For any $i\in\mathcal{N}_{S}$ and any $j\in\mathcal{N}_{S}\cup\mathcal{N}_{o}\setminus\{i\}$, by using \eqref{eq.def.p_i.kinematics} and \eqref{eq.multi.lyapunov}, we have
\begin{align}
&|V_{ij}(t)|  = \mu\left(\left|\tilde{p}_{ij}(T_S) +\int^{t}_{T_S}(v_i(\tau)-v_j(\tau)) d\tau \right|-D_s\right) \notag \\
%&\le \mu\left(|\tilde{p}_{ij}(T_S)| - \left|\int^{t}_{T_S}\tilde{v}_i(\tau)-\tilde{v}_j(\tau) d\tau \right|-D_s\right) \notag \\
&\le \mu\left(\mu^{-1}(V_{ij}(T_S)) - \int^{t}_{T_S}|\tilde{v}_i(\tau)|+|\tilde{v}_j(\tau)| d\tau\right) \notag \\
&\le \mu\left(\mu^{-1}(V_{ij}(T_S)) - c_{\tilde{v}}\lambda^{-1}(|z_i(T_S)|+|z_j(T_S)|)\right)
\end{align}
for all $t\ge T_S$. We use \eqref{eq.multi.referencetrackingerror} for the first inequality and use \eqref{eq.asmp.vtilde_i.IOpS} for the last inequality above. It is a direct consequence that
\begin{align}
|V_S(t)| \le \mu\left( \mu^{-1}(V_S(T_S)) - 2c_{\tilde{v}}\lambda^{-1} |z_m(T_S)| \right)\label{eq.the.VS.bound1}
\end{align}
for all $t\ge T_S$. This shows the boundedness of $V_S$.

{\bfseries (B2) Boundedness of $\tilde{v}_R$ and $V_R$.} Properties \eqref{eq.proposition.multi.lyapunov.ISpS} and \eqref{eq.solution.derivative.boundedness} imply that
\begin{align}
&|\tilde{v}_R(t)|\le\max \notag \\
&\left\{\begin{array}{l}
\beta_{\tilde{v}}^V(|z_m(T_S)|, t-T_S) + \gamma^{V}_{\tilde{v}}(\| V_R\|_{[T_S, t)})+d^{V}_{\tilde{v}},
\\
\gamma^{\tilde{v}}_{\tilde{v}}(\|\tilde{v}_R\|_{[T_S, t)})
\end{array}
\right\},\label{eq.the.case2.tildevR}\\
&|V_R(t)| \le \beta_V(V_R(T_S), t-T_S) + d^{\tilde{v}}_{V} \notag \\
&~~~~~ + \gamma^{\tilde{v}}_{V}(\max\{\|\tilde{v}_R\|_{[T_S, t)}, \|\tilde{v}_S\|_{[T_S, t)}\}) \label{eq.the.case2.VR}
\end{align}
for all $t\ge T_S$.
From \eqref{eq.the.vtildes.bound1}, it can be easily verified that 
\begin{align}
\|\tilde{v}_S\|_{[T_S, t)} \le c_{\tilde{v}}|z_m(T_S)|. \label{eq.the.vtildes.bound2}
\end{align}
By combining \eqref{eq.the.case2.VR} and \eqref{eq.the.vtildes.bound2}, we have 
\begin{align}
&|V_R(t)|
\le \beta_V(V_R(T_S), t-T_S) + \gamma^{\tilde{v}}_{V}(\|\tilde{v}_R\|_{[T_S, t)})\notag \\
&~~~~ + \gamma^{\tilde{v}}_{V}(c_{\tilde{v}}|z_m(T_S)|) + d^{\tilde{v}}_{V}\notag \\
&~= \beta_V(V_R(T_S), t-T_S) + \gamma^{\tilde{v}}_{V}(\|\tilde{v}_R\|_{[T_S, t)}) + d^{\tilde{v}}_{V_k}\label{eq.the.case2.VR2}
\end{align}
for all $t\ge T_S$.

Properties \eqref{eq.the.case2.tildevR} and \eqref{eq.the.case2.VR2} result in an interconnection between $\tilde{v}_R$ and $V_R$, with a structure quite similar with the one between $\tilde{v}_m$ and $V_m$ shown in Figure \ref{figure.Small gain diagram 2}.
The boundedness of $\tilde{v}_R$ and $V_R$ can be proved following the small-gain analysis as for Case (A). The boundedness of $V_m$, $\tilde{v}_m$ and $z_m$ can be proved by combining the analysis in Cases (B1) and (B2).

Due to space limitation, a step-by-step guideline of choosing controller parameters to satisfy the conditions required in the proofs above is given in the technical report \cite{technicalreport}.

This completes the proof of Theorem \ref{theorem.smallgaincondition2}.

\section{Simulation and Experiment} 
\label{section.experiment}

In this section, we consider two safety control scenarios for quadrotors. Numerical simulations and physical experiments are used to verify our approach.

The experimental system for model identification and algorithm verification is composed of quadrotors (Crazyflie 2.0 with original onboard attitude controller), an optical motion capture system (OptiTrack), and a laptop computer running the Robot Operating System (ROS). The motion capture system measures the real-time positions and velocities of the quadrotors. Data transmission from the laptop computer to the quadrotors is based on Crazyradio PA, and that from the motion capture system to the laptop computer is through a TCP/IP ethernet.

\subsection{An Identified Model of a Quadrotor and Verification of Assumption \ref{assumption.multi.referencetracking}}
\label{subsection.identification}

The quadrotor velocity is controlled by a PID controller designed by Matlab Control System Designer and implemented by ROS.
We perform system identification for the velocity-controlled quadrotor by using Matlab System Identification Toolbox based on indoor experiment data. 

The identified model with $v^*$ as the input and the actual velocity $v$ as the output is in the form of \eqref{eq.example.linearsystem} where
\begin{align}
%A&=\left[\begin{array}{cccc}
%A_{11} & 0 & 0 & 0\\
%0 & A_{22} & 0 & 0\\
%0 & 0 & A_{33} & 0\\
%0 & 0 & 0 & A_{44}
%\end{array}\right],\\
A&=\diag\{A_{11},A_{22}, A_{33}, A_{44}\},\\
B^T&=\left[\begin{array}{cccc}
B_{11}^T & B_{21}^T & 0 & 0\\
0 & 0 & B_{32}^T & B_{42}^T
\end{array}\right],\\
C&=\left[\begin{array}{cccc}
C_{11} & 0 & C_{13} & 0\\
0 & C_{22} & 0 & C_{24}
\end{array}\right],
\end{align}
with $A_{11}=[-1.58,2.92;-2.92,-1.58]$, $A_{22}=[-2.68,7.18;-7.18,-2.68]$, $A_{33}=[-2.56,6.86;-6.86,-2.56]$, $A_{44}=[-2.14,3.71;-3.71,-2.14]$, $B_{11}^T=[1.65,0.65]$, $B_{21}^T=[1.5,0.92]$, $B_{32}^T=[1.58,0.84]$, $B_{42}^T=[1.51,-2.29]$, $C_{11}=[0.78,-1.98]$, $C_{13}=[2.13,2.41]$, $C_{22}=[-2.2,-2.82]$, $C_{24}^T=[-1.51,-0.99]$.

Choose $Q = A+A^T$ and $P=I_{8\times 8}$. Then, $PA+A^TP=-Q$. Define $V(\zeta)=\zeta^TP\zeta$ with $\zeta=Az+Bv^*$. By taking the derivative of $V(\zeta)$ along the trajectories of the system \eqref{eq.example.tranformed.linearsystem1}--\eqref{eq.example.tranformed.linearsystem2}, we have
\begin{align}
&\nabla V\dot{\zeta} = \zeta^T(A^TP+PA)\zeta+2\zeta^TPBv^{*d} \notag \\
&= -\xi |\zeta|^2 -\zeta^T(Q-\xi I)\zeta+2\zeta^TPBv^{*d}  \notag \\
&\le -\xi |\zeta|^2 -|\zeta|\left(\lambda_{\min}(Q-\xi I)\sqrt{\lambda^{-1}_{\max}(P)V} - 2|PB||v^{*d}|\right)\label{eq.iden.derivative.1}
\end{align}
where $\xi$ is a constant satisfying $0\le \xi\le\lambda_{\min}(Q)$. 

It is a direct consequence that
\begin{align}
V(\zeta)\ge \left(\frac{2\lambda^{1/2}_{\max}(P)|PB||v^{*d}|}{\lambda_{\min}(Q-\xi I)} \right)^2 \!\!\Rightarrow\! \nabla V\dot{\zeta}\le -\xi |\zeta|^2. \label{eq.iden.gain.1}
\end{align}
By directly applying the definitions of input-to-state stability (ISS) and ISS-Lyapunov function \cite{Sontag-2008}, there exists a $\beta\in \mathcal{KL}$ such that 
\begin{align}
V(\zeta(t))\le  \max \left\{\beta(|\zeta(0)|, t), \left(\frac{2\lambda^{1/2}_{\max}(P)|PB||v^{*d}|}{\lambda_{\min}(Q-\xi I)} \right)^2\right\} \label{eq.iden.gain.2}
\end{align}
for all $t\ge 0$.

By applying the norm inequality to \eqref{eq.example.tranformed.linearsystem2}, we have 
\begin{align}
|\tilde{v}|^2
&\le \lambda_{\max}(C_\zeta^TC_\zeta) |\zeta|^2 
\le \frac{\lambda_{\max}(C_\zeta^TC_\zeta)}{\lambda_{\min}(P)} V(\zeta)
\end{align}
with $C_{\zeta}=CA^{-1}$.

Properties \eqref{eq.iden.gain.1} and \eqref{eq.iden.gain.2} together imply
\begin{align}
|\tilde{v}(t)|&\le 2\sqrt{\frac{\lambda_{\max}(P)\lambda_{\max}(C_\zeta^TC_\zeta)}{\lambda_{\min}(P)}}\frac{|B^TP|}{\lambda_{\min}(Q-\xi I)}|v^{*d}|\notag\\
&~~~+ \sqrt{\frac{\lambda_{max}(C_\zeta^TC_\zeta)}{\lambda_{\min}(P)}\beta(|\zeta(0)|,t)}. \label{eq.iden.IOS}
\end{align}

Clearly, \eqref{eq.iden.IOS} is in the form of \eqref{eq.asmp.vtilde_i.IOpS}. Property \eqref{eq.asmp.z_i.ISS} can also be easily validated based on the Lyapunov formulation above. Assumption \ref{assumption.multi.referencetracking} is verified.

\subsection{Collision Avoidance in the Case of One Mobile Agent and Two Obstacles}
\label{subsection.experiment.1+2}

In this subsection, we consider the scenario in Examples \ref{example.Not Lipschitz} and \ref{example.feasible_region}. The identified quadrotor model given in Subsection \ref{subsection.identification} is used for numerical simulations.

The QP algorithm with relaxation parameter and reshaped feasible set (RPRF) defined by \eqref{eq.def.calP'_i}--\eqref{eq.def.v*_i.controller} is compared with the QP algorithm with relaxation parameter (RP) defined by \eqref{eq.def.QP.non_Lipschitz}--\eqref{eq.QP2.domian.ai}, to show the effectiveness of the proposed method.

In particular, we consider constant velocity command:
\begin{align}
v_1^c(t)&\equiv[2;-2].
\end{align}

For both of the algorithms, we choose
\begin{align}
D&=0.6,&D_s&=0.65,&\delta&=100,\\
\mu(s)&=\frac{1}{s+D_s},&V(\tilde{p})&=|\tilde{p}|^{-1},&\alpha_V(s)&=0.75s.
\end{align}
For the QP algorithm with RPRF, we also choose
\begin{align}
c_K&=1,&c_P&=2\sqrt{2}
\end{align}
and $[A_1^r]_{j,:}=[\cos(2\pi j/11),~\sin(2\pi j/11)]$ for $j=1,\ldots,n_p$ with $n_p=11$. Accordingly, $c_A=\cos(2\pi/11)$.

\begin{figure}[h!]
\centering
\includegraphics[width=0.8\linewidth]{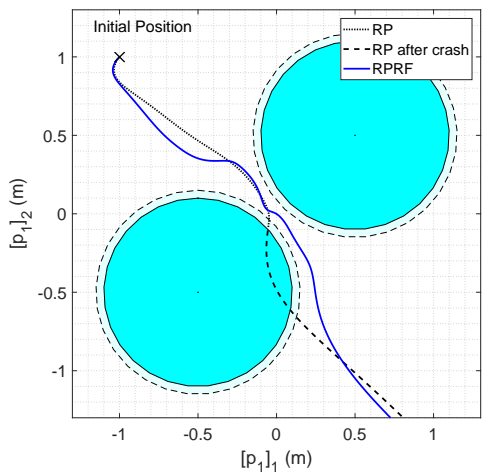}
\caption{Trajectories of the controlled mobile agent with different QP-based controllers.}
\label{fig.simulation_trajecoty}
\end{figure}

\begin{figure}[h!]
\centering
\includegraphics[width=0.95\linewidth]{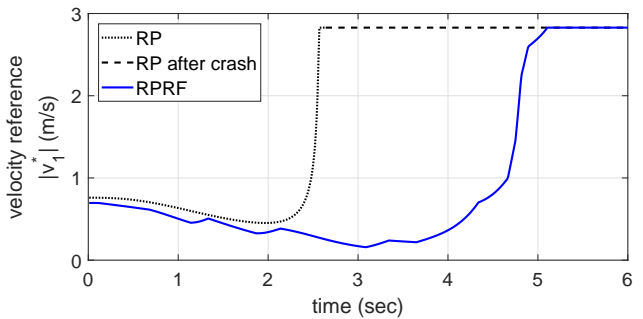}
\caption{Velocity reference signals with different QP-based controllers.}
\label{fig.simulation_velocity_reference}
\end{figure}

Figures \ref{fig.simulation_trajecoty} and \ref{fig.simulation_velocity_reference} show the trajectories of the controlled mobile agent and the velocity reference signals with the two algorithms. Due to the uncertain actuation dynamics, the velocity reference signal generated by the QP algorithm with RP leads to an unexpected response and causes collision. The QP algorithm with RPRF avoids collision.

\subsection{Collision Avoidance in the Case of Three Quadrotors}

To verify the proposed method in practice, we consider a scenario of three quadrotors swapping positions.

The primary controller is trajectory tracking. The reference trajectories of the agents are generated by sinusoidal functions:
\begin{align}
p^r_1(t)&=[-\cos(2\pi t/15);-\cos(2\pi t/15);1], \\
p^r_2(t)&=[\cos(2\pi t/15)-0.23;\cos(2\pi t/15)+0.23;1],\\
p^r_3(t)&=[\cos(2\pi t/15)+0.23;\cos(2\pi t/15)-0.23;1].
\end{align}
And the velocity commands are generated by feedforward-feedback controllers:
\begin{align}
v_i^c(t)=-1.7(p_i(t)-p_i^r(t))+\dot{p}_i^r(t)
\end{align}
for $i=1,2,3$.

For both of the algorithms, we choose
\begin{align}
D&=0.2,&D_s&=0.3,&\delta&=100,\\
\mu(s)&=\frac{1}{s+D_s},&V(\tilde{p})&=|\tilde{p}|^{-1},&\alpha_V(s)&=0.3s.
\end{align}
For the QP algorithm with RPRF, we also choose
\begin{align}
c_K&=1,&c_P&=1
\end{align}
and $[A_1^r]_{j,:}=[\cos(2\pi j/n_p),~\sin(2\pi j/n_p)]$ for $j=1,\ldots,n_p$ with $n_p=11$. Accordingly, $c_A=\cos(2\pi/11)$.

\begin{figure}[h!]
\centering
\includegraphics[width=1.0\linewidth]{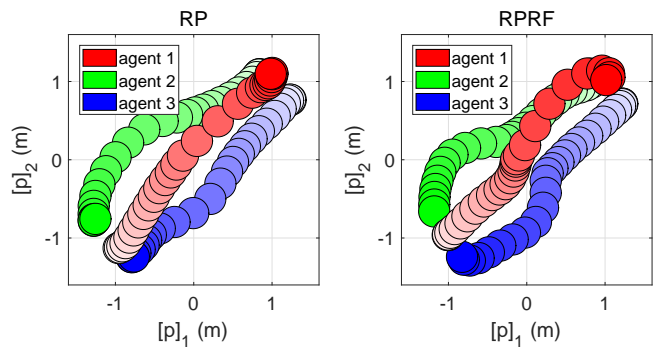}
\caption{Trajectories of the agents with different QP-based controllers. }
\label{fig:exp3uavtrajectorytracking}
\end{figure}

\begin{figure}[h!]
\centering
\includegraphics[width=1.0\linewidth]{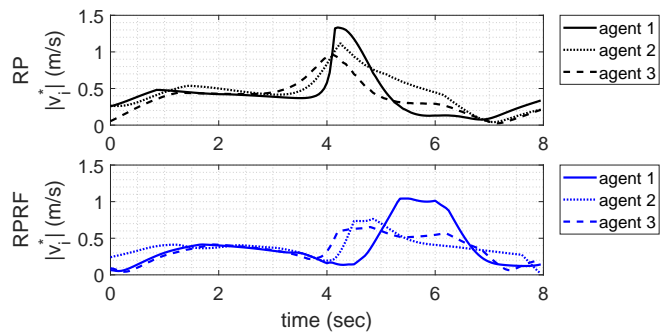}
\caption{Norms of the velocity reference signals with different QP-based controllers.}
\label{fig:exp3uavmaximalderivative}
\end{figure}

Figures \ref{fig:exp3uavtrajectorytracking} and \ref{fig:exp3uavmaximalderivative} show the trajectories of the controlled mobile agents and the norms of the velocity reference signals with the two algorithms. Although collision avoidance is achieved in both of the cases, one may recognize a sudden change of agent 1's velocity reference signal at $t=4.1$ when the QP-based algorithm without feasible-set reshaping is applied.

\section{Conclusions}
\label{section.conclusion}

This paper has developed a systematic solution to the control of safety-critical multi-agent systems subject to uncertain actuation dynamics. 
The major contribution lies in a seamless integration of a new QP-based design with reshaped feasible set and a nonlinear small-gain analysis. 
In particular, the new feasible-set reshaping technique has been proved to be quite useful for feasibility of the refined QP algorithm and Lipschitz continuity of its solution.  The nonlinear small-gain analysis takes advantage of the interconnection between the controlled nominal system and the uncertain actuation system for ensured safety.

We imagine that the new techniques are beneficial to solving multi-objective control problems for more general systems, e.g., control-affine systems and nonholonomic systems, and control systems subject to information constraints, e.g., partial-state feedback and sampled-data feedback. It is also of theoretical and practical interest to study distributed and coordinated implementations of the algorithms.

\appendices

\section{Technical Lemmas on Linear Spans and Positive Spans}
\label{appendix.lemma.positivecombination}

The following two lemmas are used to prove the implementability of the feasible-set reshaping technique proposed in this paper. One may consult \cite{Davis-AJM-1954} for basic notions of positive linear combination. Due to space limitation, the proofs of the lemmas are given in the technical report \cite{technicalreport}.

\begin{lemma}\label{lemma.linearspan}
Suppose that any $n$ of the nonzero vectors $v_1,\ldots,v_m\in\mathbb{R}^n$ with $m\geq n$ are linearly independent. For any specific $v^*\in\mathbb{R}^n$ and any positive constant $\epsilon$, one can find a vector $v_{m+1}\in\mathbb{R}^n$ such that $|v_{m+1}-v^*|\leq\epsilon$ and any $n$ of the vectors $v_1,\ldots,v_{m+1}$ are linearly independent.
\end{lemma}

\begin{lemma}\label{lemma.positivespan}
There exist $v_1,\ldots,v_r\in\mathbb{R}^n$ with $r\geq n+1$, any $n$ of which are linearly independent, to positively span $\mathbb{R}^n$.
\end{lemma}

\bibliographystyle{IEEEtran}% Include this if you use bibtex 
\bibliography{CAreferencesL} % and a bib file to produce the 

\begin{IEEEbiography}[{\includegraphics[width=1in,height=1.25in,clip,keepaspectratio]{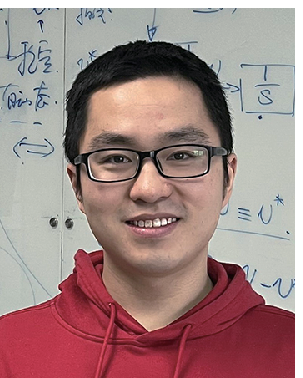}}]{Si Wu} received the B.E. degree in Automation from Henan Polytechnic University, China, in 2017, and the M.E. degree in Control Engineering from Northeastern University, China, in 2020. He is pursuing his Ph.D. degree in Control Science and Engineering at Northeastern University. His research interest focuses on nonlinear control theory and applications to vehicles and robotic systems. He was a leading team member winning the IMAV 2019 outdoor competition.
\end{IEEEbiography}

\begin{IEEEbiography}[{\includegraphics[width=1in,height=1.25in,clip,keepaspectratio]{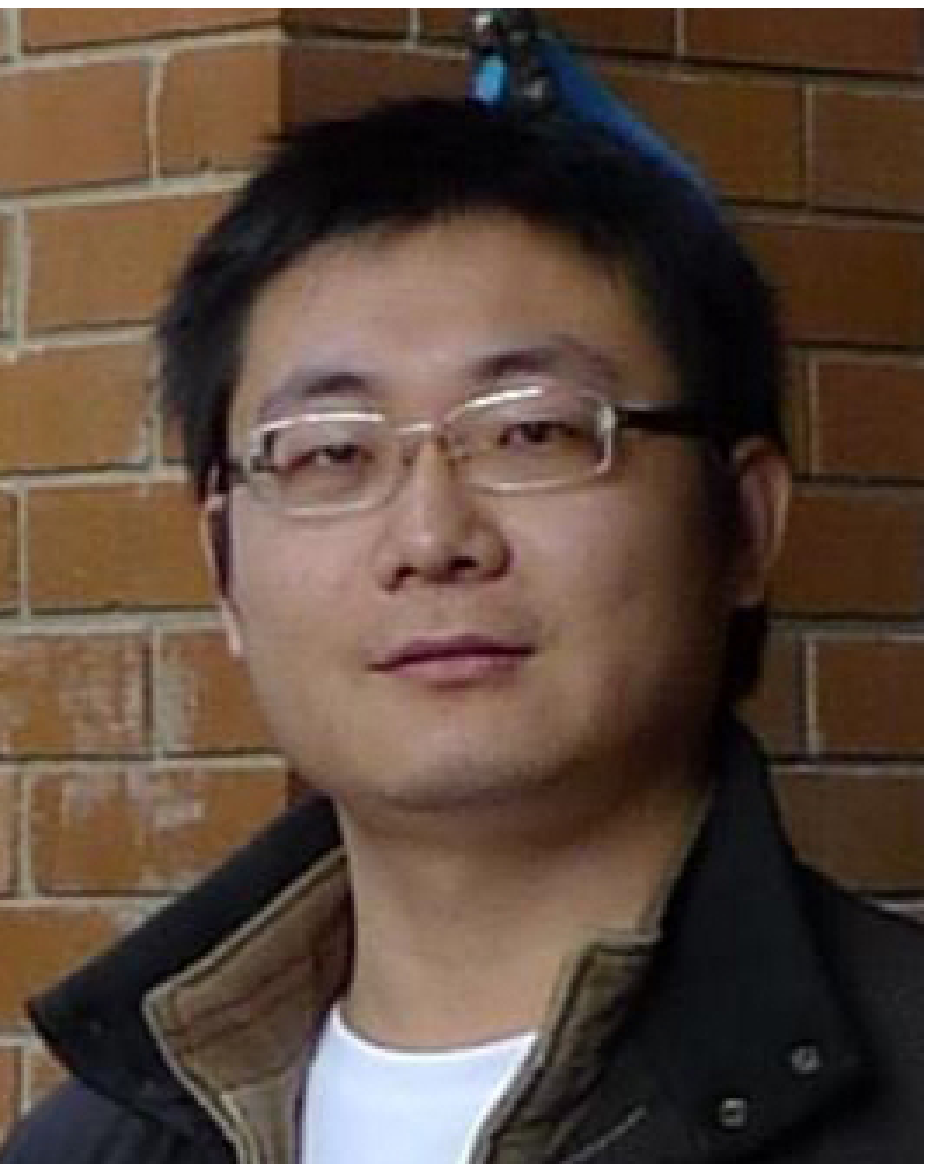}}]{Tengfei Liu} received the B.E. degree in automation, in 2005, the M.E. degree in control theory and control engineering, in 2007,
both from South China University of Technology, Guangzhou, China, and the Ph.D. degree in engineering from RSISE, the Australian National University, Canberra, Australia, in 2011. From 2011 to 2013, he was a Postdoc with faculty fellowship at Polytechnic Institute of New York University. Since 2014, he has been a Faculty Member with Northeastern University, Shenyang, China.

His research interests include stability and control of interconnected nonlinear systems.
\end{IEEEbiography}

\begin{IEEEbiography}[{\includegraphics[width=1in,height=1.25in,clip,keepaspectratio]{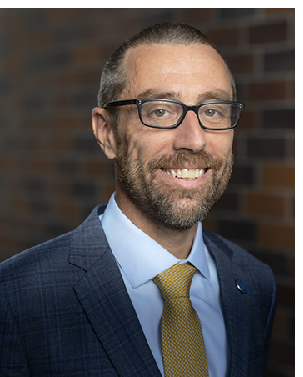}}]{Magnus Egerstedt} is the Dean of Engineering and a Professor in the Department of Electrical Engineering and Computer Science at the University of California, Irvine. Prior to joining UCI, Egerstedt was on the faculty at the Georgia Institute of Technology. He received the M.S. degree in Engineering Physics and the Ph.D. degree in Applied Mathematics from the Royal Institute of Technology, Stockholm, Sweden, the B.A. degree in Philosophy from Stockholm University, and was a Postdoctoral Scholar at Harvard University. Dr. Egerstedt conducts research in the areas of control theory and robotics, with particular focus on control and coordination of multi-robot systems. Magnus Egerstedt is a Fellow of IEEE and IFAC, and is a Foreign member of the Royal Swedish Academy of Engineering Science. He has received a number of teaching and research awards, including the Ragazzini Award, the O. Hugo Schuck Best Paper Award, the Outstanding Doctoral Advisor Award and the HKN Outstanding Teacher Award from Georgia Tech, and the Alumni of the Year Award from the Royal Institute of Technology.
\end{IEEEbiography}

\begin{IEEEbiography}[{\includegraphics[width=1in,height=1.25in,clip,keepaspectratio]{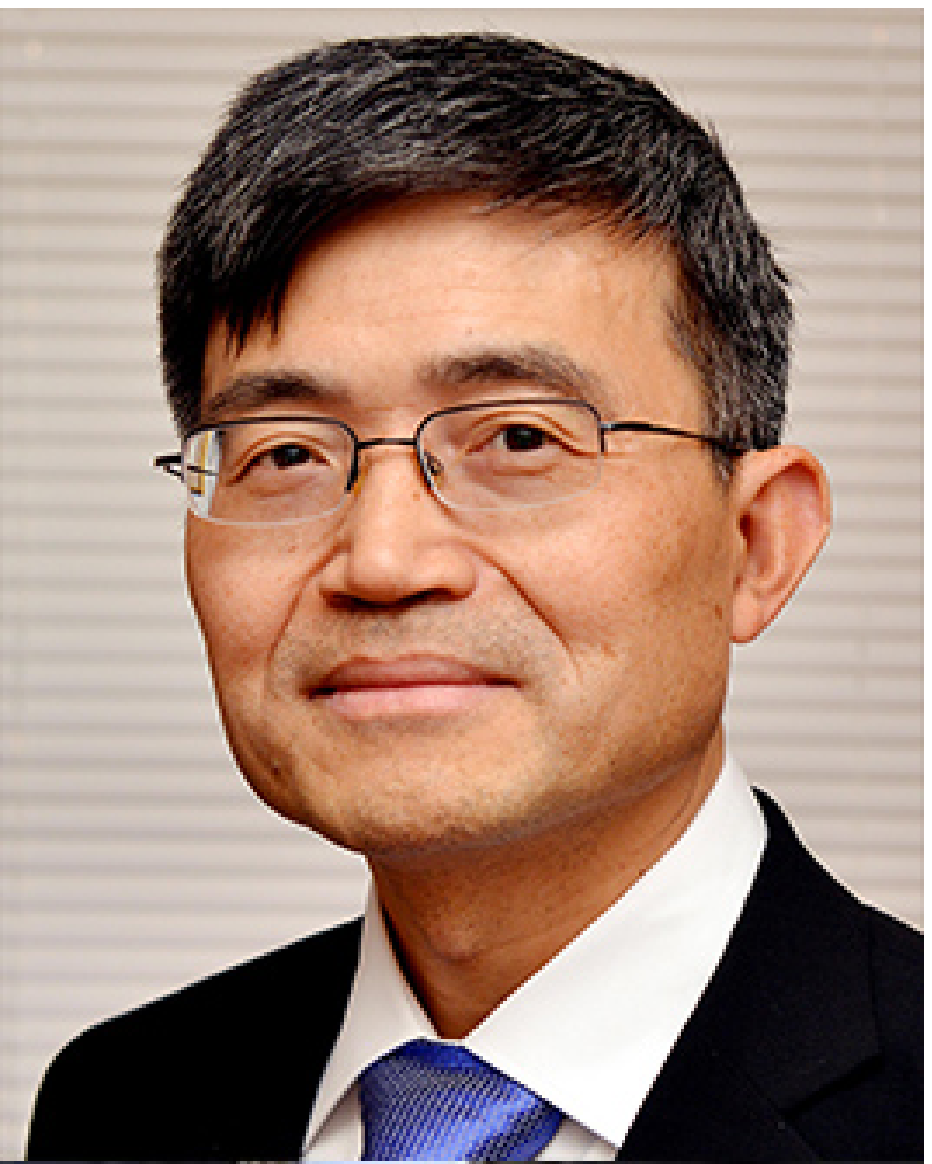}}]{Zhong-Ping Jiang} received the M.Sc. degree in statistics from the University of Paris XI, France, in 1989, and the Ph.D. degree in automatic control and mathematics from the Ecole des Mines de Paris (now, called ParisTech-Mines), France, in 1993, under the direction of Prof. Laurent Praly.

Currently, he is a Professor of Electrical and Computer Engineering at the Tandon School of Engineering, New York University. His main research interests include stability theory, robust/adaptive/distributed nonlinear control, robust adaptive dynamic programming, reinforcement learning and their applications to information, mechanical and biological systems. 

He has served as Deputy Editor-in-Chief, Senior Editor and Associate Editor for numerous journals. Prof. Jiang is a Fellow of the IEEE, a Fellow of the IFAC, a Fellow of the CAA and is among the Clarivate Analytics Highly Cited Researchers. In 2021, he is elected as a foreign member of the Academy of Europe.

\end{IEEEbiography}

\end{document}